\documentclass[twocolumn,showpacs,nofootinbib,preprintnumbers,amsmath,amssymb,showkeys]{revtex4}

\usepackage{epsfig}
\usepackage{dsfont}
\usepackage{color}
\usepackage{graphicx}
\usepackage{dcolumn}
\usepackage{bm}

\def\slashii#1{\setbox0=\hbox{$#1$}             
   \dimen0=\wd0                                 
   \setbox1=\hbox{\sl/} \dimen1=\wd1            
   \ifdim\dimen0>\dimen1                        
      \rlap{\hbox to \dimen0{\hfil\sl/\hfil}}   
      #1                                        
   \else                                        
      \rlap{\hbox to \dimen1{\hfil$#1$\hfil}}   
      \hbox{\sl/}                               
   \fi}                                         %
\def\slashiii#1{\setbox0=\hbox{$#1$}#1\hskip-\wd0\hbox to\wd0{\hss\sl/\/\hss}}
\newcommand{\beq}{\begin{equation}}
\newcommand{\eeq}{\end{equation}}
\newcommand{\bea}{\begin{eqnarray}}
\newcommand{\eea}{\end{eqnarray}}
\newcommand{\nn}{\nonumber \\}

\newcommand\eqn[1]{(\ref{#1})}      
\newcommand\Eqn[1]{Eq.~(\ref{#1})}  
\newcommand\Fig[1]{Fig.~\ref{#1}}  
\newcommand\Sec[1]{Sec.~\ref{#1}}  

\begin{document}


\title{Quantum scalar fields in de Sitter space from the nonperturbative \\renormalization group}

\author{Maxime Guilleux}

\author{Julien Serreau}

\affiliation{APC, AstroParticule et Cosmologie, Universit\'e Paris Diderot, CNRS/IN2P3, CEA/Irfu, Observatoire de Paris, Sorbonne Paris Cit\'e \\
10, rue Alice Domon et L\'eonie Duquet, 75205 Paris Cedex 13, France
}

\date{\today}

\begin{abstract}

We investigate scalar field theories in de Sitter space by means of nonperturbative renormalization group techniques. We compute the functional flow equation for the effective potential of O($N$) theories in the local potential approximation and we study the onset of curvature-induced effects as quantum fluctuations are progressively integrated out from subhorizon to superhorizon scales. This results in a dimensional reduction of the original action to an effective zero-dimensional Euclidean theory. We show that the latter is equivalent both to the late-time equilibrium state of the stochastic approach of Starobinsky and Yokoyama and to the effective theory for the zero mode on Euclidean de Sitter space. We investigate the immediate consequences of this dimensional reduction: symmetry restoration and dynamical mass generation.

 \end{abstract}

\maketitle

\section{Introduction}

Space-time curvature can have important consequences on the dynamics of quantum fields. Prominent examples are the spontaneous Hawking/Unruh radiation from (analog) black holes \cite{Hawking:1974sw,Unruh:1976db,Brout:1995rd} and the  amplification of cosmological perturbations during the inflation era \cite{Mukhanov:1990me}. Other nontrivial effects include the possibility of gravitationally induced phase transitions in the early Universe \cite{Shore:1979as,Buchbinder1992,Elizalde:1993ee}, the decay of massive particles into themselves \cite{Bros:2006gs,Jatkar:2011ju}, the generation of a nonvanishing photon mass \cite{Prokopec:2002jn,Prokopec:2003tm}, or the phenomenon of symmetry restoration through gravitationally enhanced quantum fluctuations \cite{Ratra:1984yq,Mazzitelli:1988ib,Janssen:2009pb,Serreau:2011fu}, just to name a few. More generally, the study of radiative corrections to quantum field dynamics in nontrivial gravitational backgrounds is the subject of intense investigations; see, e.g., \cite{Anderson:1985hz,Elizalde:1994ds,Tsamis:1996qk,Onemli:2002hr,Brunier:2004sb,Boyanovsky:2005px,Sloth:2006az,Seery:2007we,Urakawa:2008rb,Marolf:2010nz,Hollands:2010pr,Higuchi:2010xt,Tanaka:2013caa,Onemli:2013gya,Herranen:2013raa,Herranen:2015aja}.

De Sitter space-time plays a particular role in this context, first, because it is maximally symmetric and, second, because of its direct relevance to inflationary physics and to the recent acceleration of the Universe \cite{Peiris:2003ff,Perlmutter:1998np}. For free scalar fields with a small mass in units of the space-time curvature, the de Sitter kinematics results in large quantum fluctuations on superhorizon scales, with an almost scale invariant power spectrum.
This is at the very origin of the success of inflationary cosmology in predicting the spectrum of primordial density fluctuations \cite{Parentani:2004ta,Langlois:2010xc}. However, this is also responsible for infrared and secular divergences in perturbative calculations of quantum (loop)  corrections to scalar field dynamics in de Sitter space \cite{Tsamis:2005hd,Weinberg:2005vy}. In fact, gravitationally enhanced quantum fluctuations on superhorizon scales lead to genuine nonperturbative effects \cite{Tanaka:2013caa,Serreau:2013koa}.

Specific techniques beyond standard perturbation theory have been developed to capture the dynamics of the relevant modes. This ranges from the effective stochastic approach put forward in Ref.~\cite{Starobinsky:1994bd} to various quantum field theoretical methods suitably adapted to de Sitter space; see Refs.~\cite{vanderMeulen:2007ah,Burgess:2009bs,Rajaraman:2010xd,Beneke:2012kn,Akhmedov:2011pj,Garbrecht:2011gu,Boyanovsky:2012qs,Parentani:2012tx,Gautier:2013aoa,Youssef:2013by,Boyanovsky:2015tba} for a (non exhaustive) list of examples. In particular, such methods allow one to study how an interacting scalar theory cures its infrared and secular problems, e.g., with the dynamical generation of a nonzero mass.

Nonperturbative renormalization group (NPRG) methods are particularly adapted for dealing with nontrivial infrared physics in many instances, from critical phenomena in statistical physics to the long distance dynamics of non-Abelian gauge fields \cite{Berges:2000ew,Delamotte:2007pf,Gies:2006wv,Reuter:1996cp}. Such techniques have recently been formulated in de Sitter space-time\footnote{See also \cite{Gies:2013dca,Shapiro:2015ova,Benedetti:2014gja} for other recent applications in curved spaces.} in Refs.~\cite{Kaya:2013bga,Serreau:2013eoa}, where they have been used to study the renormalization group (RG) flow of O($N$) scalar field theories at superhorizon scales. A remarkable observation is that, thanks to gravitationally enhanced infrared fluctuations, the RG flow gets effectively dimensionally reduced to that of a zero-dimensional Euclidean field theory \cite{Serreau:2013eoa}. This has various consequences, such as, e.g., the radiative restoration of spontaneously broken symmetries in any space-time dimension.\footnote{ The phenomenon of radiative symmetry restoration for O($N$) scalar theories in de Sitter space-time has been firmly established both for the case of a continuous Abelian symmetry $N=2$ \cite{Ratra:1984yq} and in the limit $N\to\infty$ \cite{Mazzitelli:1988ib,Serreau:2011fu}, where exact results can be obtained. It has been convincingly demonstrated for generic values of $N$ using the stochastic approach \cite{Lazzari:2013boa} and NPRG techniques \cite{Serreau:2013eoa}. It is to be mentioned that some studies \cite{Prokopec:2011ms,Arai:2011dd,Nacir:2013xca,Boyanovsky:2012nd} find a possible (de Sitter invariant) broken symmetry phase for finite $N$. However, for continuous symmetries ($N\ge2$), the Goldstone modes acquire a nonzero mass, which is rather unphysical. We believe these are artifacts of the various approximation schemes employed in these works. For instance, the Hartree approximation used in Refs.~\cite{Prokopec:2011ms,Arai:2011dd,Nacir:2013xca} is known to produce similar spurious solutions in flat space-time at finite temperature \cite{Reinosa:2011ut}.} 

In the present work, we extend the NPRG study of Ref.~\cite{Serreau:2013eoa} and investigate the flow of the effective potential of O($N$) theories from the flat space-time (Minkowski) regime at subhorizon scales to the regime of superhorizon momenta, with fully developed curvature effects. Using the so-called local potential approximation (LPA), we study in detail the onset of gravitational effects at the horizon scale.

The phenomenon of effective dimensional reduction mentioned above allows us to establish a direct relation between the present NPRG approach and the stochastic effective theory of Starobinsky and Yokoyama \cite{Starobinsky:1994bd}. In particular, we show that the effective zero-dimensional field theory which results from integrating out the superhorizon degrees of freedom is equivalent to the late-time equilibrium state of the stochastic description. We also discuss our approach in relation with recent studies on Euclidean de Sitter space \cite{Rajaraman:2010xd,Beneke:2012kn,Benedetti:2014gja}. We show that the dimensionally reduced theory in (Lorentzian) de Sitter space-time at superhorizon scales is equivalent to the effective theory for the zero mode on the compact Euclidean de Sitter space. This provides a direct link between Euclidean de Sitter calculations and the stochastic approach. This also adds to the quantum field theoretical foundations of the latter \cite{Tsamis:2005hd,Miao:2006pn,Prokopec:2007ak,Rigopoulos:2013exa,Garbrecht:2013coa,Garbrecht:2014dca,Onemli:2015pma}.

Finally, we discuss the consequences of the dimensional reduction in the infrared by explicitly solving the functional RG flow equation for the effective potential in various situations of interest. We show that, in the cases of theories which would be either critical or in the broken phase in Minkowski space, the curvature-induced effects lead to symmetry restoration and dynamical mass generation. This is nicely illustrated in the limit $N\to\infty$, where we can solve the full functional flow equation analytically in the infrared. We argue that the large-$N$ limit actually gives the correct qualitative picture for arbitrary $N$ and, using the equivalent zero-dimensional field theory, we compute the effective mass and coupling parameters in the deep infrared. We recover and extend known results of the stochastic approach.

The paper is organized as follows. Section \ref{sec:setup} briefly reviews the NPRG setup in de Sitter space-time and the derivation of the flow equation for the effective potential in the LPA. We discuss the various regimes of interest and the phenomenon of dimensional reduction in \Sec{sec:onset}, where we also establish the relations with the stochastic approach and with Euclidean de Sitter space respectively. Explicit solutions of the functional flow equation are discussed in the large-$N$ limit and at finite $N$ in Secs.~\ref{sec:largeN} and \ref{sec:finiteN}. Some technical details are presented in the Appendices.

\section{General setup}
\label{sec:setup}

We consider a scalar field theory with $O(N)$ symmetry on the expanding Poincar\'e patch of a de Sitter space-time with Lorentzian signature in $D=d+1$ dimensions. In terms of the conformal time $-\infty<\eta<0$ and of comoving spatial coordinates ${\bf X}$, the line element reads
\begin{equation}
 ds^2 = a^2(\eta) \left(-d\eta^2 +d{\bf X}^2\right) \quad{\rm with}\quad a(\eta) = -1/\eta,
\end{equation}
in units where the expansion rate $a'/a^2=1$. The classical action reads
\begin{equation}
S[\varphi] = -\int_x \left\{\frac{1}{2} \partial_\mu\varphi_a\partial^\mu\varphi_a+V(\varphi)  \right\},
\end{equation}
where  $\int_x = \int d^Dx\sqrt{-g(x)}=\int d\eta a^D(\eta)\int d^dX $ is the invariant integration measure, with $g(x)$ the determinant of the metric tensor, the potential $V(\varphi)$ is a function of the $O(N)$ invariant $\varphi_a\varphi_a$, and a summation over repeated space-time or $O(N)$ indices $a=1,\ldots,N$ is understood. Note that the potential $V(\varphi)$ includes possible couplings to the (constant) space-time curvature.

Correlation functions for the scalar field can be computed by means of path integral techniques with weight $\exp(iS)$. In order to keep the large contributions from long wavelength quantum fluctuations under control, one introduces the modified action $S_\kappa = S + \Delta S_\kappa $, with
\begin{align}
\label{eq:ir_reg}
\Delta S_\kappa[\varphi] = \frac{1}{2}\int_{x,x'} \varphi_a(x) R_{\kappa}(x,x') \varphi_a(x'),
\end{align}
where the infrared regulator $R_{\kappa}$ acts as a large mass term for (quantum) fluctuations on sizes larger than $1/\kappa$ and essentially vanishes for short wavelength modes, thereby suppressing the contribution from the former to the path integral.\footnote{The distinction between long and short wavelength modes is ambiguous in spaces with Lorentzian signature. Here, we make this distinction on (Euclidean) constant-time hypersurfaces; see below.} From the generating functional
\begin{equation}
e^{iW_\kappa[J]} = \int {\cal D} \varphi ~\exp \bigg ( iS_\kappa[\varphi] + i\!\!\int_x J_a \varphi_a \bigg), \label{functional}
\end{equation}
one defines the regulated effective action
\begin{equation}
\label{eq:averageaction}
\Gamma_\kappa[\phi] = W_\kappa[J] -\int_x J_a \phi_a - \Delta S_\kappa[\phi],
\end{equation}
where $J$ and $\phi$ are related through $\delta W_\kappa[J]/\delta J=\phi$. The functional \eqn{eq:averageaction} smoothly interpolates between the classical action at the ultraviolet scale\footnote{The ultraviolet scale is implicitly assumed to be much larger than any other scale in the problem, e.g., $\Lambda^2\gg1,V''(\varphi)$.} $\kappa=\Lambda$, that is, $\Gamma_\Lambda[\phi]=S[\phi]$, and the standard effective action---the generating functional of one-particle-irreducible vertex functions---at the scale $\kappa=0$, where all quantum fluctuations have been integrated out, namely, $\Gamma_{\kappa=0}[\phi]=\Gamma[\phi]$. It can roughly be seen as an effective action for the physics at a scale $\kappa$. The dependence on $\kappa$ is controlled by the Wetterich equation \cite{Wetterich:1992yh}
\begin{equation}
\dot \Gamma_\kappa= \frac{i}{2}{\rm Tr}\left\{\dot R_{\kappa}\left(\Gamma_\kappa^{(2)}+R_\kappa\right)^{-1}\right\},\label{flot}
\end{equation}
where the dot denotes a derivative with respect to the RG time $\ln\kappa$ and $\Gamma_{\kappa,ab}^{(2)}(x,y)= [g(x)g(y)]^{-1/2}\delta^2\Gamma_\kappa[\phi]/\delta\phi_a(x)\delta\phi_b(y)$ is the covariant two-point vertex function. Here, the functional inversion, matrix product, and trace ${\rm Tr}$ involve both space-time variables and $O(N)$ indices.\footnote{A technical comment is in order. The calculation of the correlation functions of interest here can be conveniently formulated as an initial-value problem, where initial conditions corresponding to the quantum state of interest are specified in the infinite past (see below). This is the typical setup of a nonequilibrium problem \cite{Berges:2004yj}. In that case, standard functional techniques can be generalized by formulating the theory on Schwinger's closed time contour ${\cal C}$ \cite{Schwinger:1960qe}. In the present context, this amounts to the replacement $\int d\eta\to\int_{\cal C}d\eta$ and $\delta(\eta-\eta')\to\delta_{\cal C}(\eta-\eta')$; see, e.g., Ref.~\cite{Parentani:2012tx} for details. Discussions of NPRG methods for nonequilibrium systems can be found in Refs.~\cite{Gasenzer:2008zz,Canet:2011wf,Berges:2012ty}.} 

The functional partial differential equation \eqn{flot} cannot be solved in a closed form in general. In the present work, we are interested in the flow of the effective potential $V_\kappa(\phi)$ defined as  $\Gamma_\kappa[\phi={\rm const.}]=-\int_x V_\kappa(\phi)$. To this purpose, we evaluate \Eqn{flot} at constant field and employ the local potential ansatz (LPA)
\begin{equation}
\Gamma_\kappa^{\rm LPA} [\phi] = -\int_x \left\{ \frac{1}{2}\partial_\mu \phi_a\partial^\mu \phi_a +V_\kappa(\phi)\right\} \label{LPA}
\end{equation}
to compute the right-hand side of the equation. This is motivated by the expectation that terms with higher powers of field derivatives should be suppressed in the physically relevant regime $\kappa\lesssim1$. The LPA further neglects a possible field-dependent renormalization factor of the derivative term. It is the simplest nontrivial ansatz which incorporates the full field dependence of the effective potential. Notice that one has $V_{\Lambda}(\phi)\approx  V(\phi)$ at the ultraviolet scale $\kappa=\Lambda$.

Following \cite{Kaya:2013bga,Serreau:2013eoa}, we choose an infrared regulator of the form 
\begin{align}
 R_\kappa(x,x')&=-\frac{\delta(\eta-\eta')}{a^D(\eta)}\int\frac{d^dK}{(2\pi)^d}e^{i{\bf K}\cdot({\bf X}-{\bf X}')} R_\kappa(-K\eta)\nn
 \label{eq:regreg}
 &=-\delta(t-t')\int\frac{d^dp}{(2\pi)^d}e^{i{\bf p}\cdot({\bf x}-{\bf x}')} R_\kappa(p),
\end{align}
where, in the second line, we introduced the cosmological time $t=-\ln(-\eta)$ as well as the physical coordinates and momentum variables, ${\bf x}=a(\eta){\bf X}$ and ${\bf p}={\bf K}/a(\eta)$. When plugged in \Eqn{eq:ir_reg}, one checks that this indeed leads to a momentum-dependent mass term. An important remark is that this only regulates spatial momenta and thus breaks the local Lorentz symmetry of de Sitter space-time. The difficulty of choosing a fully invariant regulator is related to the fact that the distinction between high and low momentum modes is ambiguous in a space with Lorentzian signature. We emphasize though that it is important to regulate physical momenta $p=-K\eta$ in order to keep as much as possible of de Sitter symmetries \cite{Serreau:2013eoa}. In particular, this guarantees that the affine subgroup of the de Sitter group is left unbroken \cite{Busch:2012ne,Adamek:2013vw} and this leads to a consistent\footnote{For instance, a regulator on comoving momenta leads to inconsistencies such as the fact that one cannot factor out the volume factor $\int_x$ on both sides of \Eqn{flot}; see Ref.~\cite{Kaya:2013bga}.} truncation of both sides of the flow equation \eqn{flot}.

\begin{figure}
\begin{center}
\includegraphics[width=0.4\textwidth]{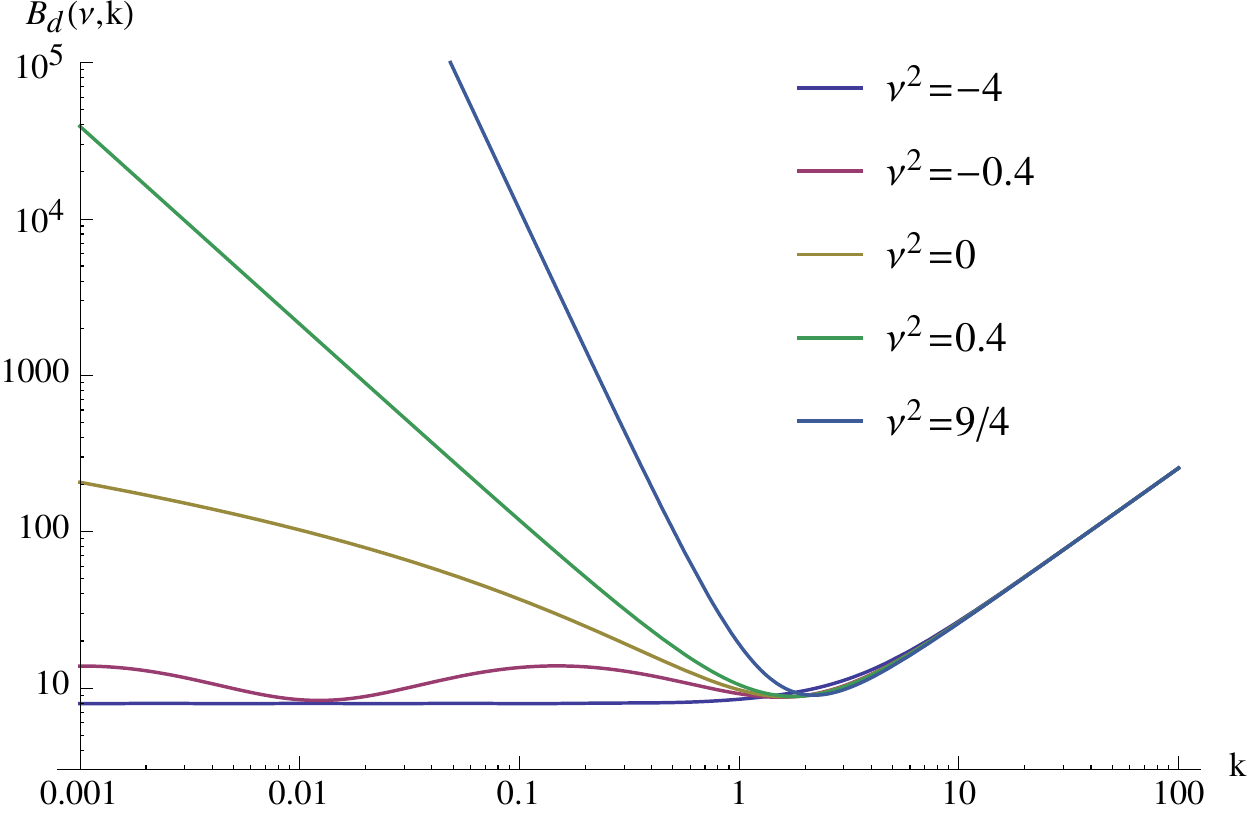}
\caption{The function $B_d(\nu,\kappa)$ [see \Eqn{nu_inv}] in $D=3+1$ dimensions versus $\kappa$ for various (real and imaginary) values of $\nu$.  In the UV regime $\kappa\gtrsim1$ the function $B_d(\nu,\kappa)\sim\kappa$, which reproduces the Minkowski beta function for the potential. Imaginary values of $\nu$ correspond to regions of field space where the curvature of the potential $V_\kappa''>d^2/4$. In that case, the function $B_d(\nu,\kappa)$ shows a bounded oscillatory behavior for $\kappa\lesssim1$ and it is essentially constant for large field curvatures, $V_\kappa''\gg d^2/4$. For $\nu=0$, this turns into a logarithmic behavior, which reflects the gravitational enhancement of superhorizon fluctuations. Finally, real positive values of $\nu$ correspond to regions of field space where the curvature of the potential $V_\kappa''<d^2/4$ and are most sensitive to space-time curvature effects. The logarithmic enhancement is turned into a power law $\kappa^{-2\nu}$.}\label{eq:Bd}
\label{Bfig}
\end{center}
\end{figure}

With these choices, the flow equation for the potential takes the following form, in the case $N=1$ and keeping the field dependence implicit,  
\begin{equation}
\dot V_\kappa = \frac{1}{2 } \int \frac{d^d p}{(2\pi)^{d}} \dot{ R}_\kappa (p) \frac{| u_\kappa(p)|^2 }{p} \label{pot_flow0},
\end{equation}
where the mode function $u_\kappa(p)$ satisfies the evolution equation\footnote{In general cosmological space-times, the mode function depends separately on the comoving momentum $K$ and the conformal time $\eta$. The symmetries of the de Sitter space-time---in fact the affine subgroup \cite{Busch:2012ne,Adamek:2013vw}---constrain these dependences to be tight together by the gravitational redshift. The mode function is a nontrivial function of the physical momentum $p=-K\eta$ only. The time-evolution equation can be traded for a (physical) momentum evolution equation; see Refs.~\cite{Busch:2012ne,Adamek:2013vw,Parentani:2012tx} for details.}
\beq
 \left(\partial_p^2+1-\frac{\nu_\kappa^2-R_\kappa(p)-{1\over4}}{p^2}\right)u_\kappa(p)=0,
\eeq
with appropriate initial conditions, where
\beq
\label{eq:nu}
 \nu_\kappa=\sqrt{\frac{d^2}{4}-V_\kappa''}.
\eeq
For the simple Litim regulator \cite{Litim:2001up}
\begin{equation}
  R_\kappa(p) = (\kappa^2 - p^2)\theta(\kappa^2 - p^2) \label{reg}
\end{equation}
and demanding the Bunch-Davies \cite{Bunch:1978yq} vacuum conditions at large momentum (which reproduce the Minkowski vacuum for deep subhorizon modes), the solution reads
\begin{align}
u_\kappa (p) &= \sqrt{\frac{\pi p}{4}} e^{i\varphi_\kappa}\!\left[c_\kappa^+ \left ( \frac{p}{\kappa} \right)^{\!\bar \nu_\kappa} + c_\kappa^- \left ( \frac{\kappa}{p} \right)^{\!\bar \nu_\kappa}\right]\,\,\,{\rm for }\,\,\,p\le\kappa\nn
u_\kappa (p) &= \sqrt{\frac{\pi p}{4}} e^{i\varphi_\kappa}H_{\nu_\kappa}(p)\quad{\rm for }\quad p\ge\kappa,
\end{align}
where $\varphi_\kappa=\frac{\pi}{2}(\nu_\kappa + {1/2})$,  $\bar \nu_\kappa^2 = \nu_\kappa^2 - \kappa^2$, $H_\nu(p)$ is the Hankel function of the first kind, and where the 
coefficients 
\begin{equation}
 c_\kappa^\pm = \frac{1}{2} \left[ H_{\nu_\kappa}(\kappa) \pm \frac{\kappa}{\bar \nu_\kappa}H_{\nu_\kappa}'(\kappa) \right ]
\end{equation}
ensure the continuity of $u_\kappa(p)$ and of its first derivative at $p=\kappa$. The momentum integral in \Eqn{pot_flow0} can be computed explicitly. We obtain the functional beta function for the potential as
\beq
\dot V_\kappa \equiv \beta(V_\kappa'',\kappa) =   \frac{C_d \kappa^{d+2}}{\kappa^2 + V_\kappa''}B_d(\nu_\kappa,\kappa), \label{pot_flow}
\eeq
where $C_d = \pi \Omega_d /[16 d(2\pi)^d]$, with $\Omega_d=2\pi^{d/2}/\Gamma(d/2)$, and where we have defined the function\footnote{ A. Kaya has informed us that the beta function published in Ref.~\cite{Kaya:2013bga} contains two typos: $3+3n\to3+2n$ and $9+6n-2\alpha^2\to9+6n+2\alpha^2$. Our results agree once these typos are corrected.} (see \Fig{eq:Bd})
\begin{align}
\label{nu_inv} 
B_d(\nu,\kappa) &= e^{- \pi  {\rm Im}(\nu)} \bigg\{ \left(d^2 - 2 \nu^2+ 2 \kappa^2\right) \big|H_\nu(\kappa)\big|^2 \nn
& + 2\kappa^2 \big|H_\nu'(\kappa)\big|^2 - 2 d \kappa\, {\rm Re}\big[H_\nu^*(\kappa) H_\nu' (\kappa)\big] \bigg\}.
\end{align}

The generalization to the case $N>1$ is straightforward. Defining
\begin{equation}
V_\kappa(\phi) = NU_\kappa(\rho) \quad \text{with} \quad \rho=\frac{\phi_a\phi_a}{2N},
\end{equation}
we obtain the functional flow equation
\beq
\label{eq:flot_N}
N \dot U_\kappa =  \beta\left(m^2_{l,\kappa} ,\kappa\right)+(N-1)\,\beta\left(m^2_{t,\kappa},\kappa\right),
\eeq
with the local curvatures in the longitudinal and transverse directions in field space 
\beq
\label{eq:flot_N2}
 m^2_{l,\kappa}(\rho) =U_\kappa'(\rho) + 2\rho U_\kappa''(\rho) \,\,\,\,{\rm and}\,\,\,\, m^2_{t,\kappa}(\rho) = U_\kappa'(\rho).
\eeq

\section{From subhorizon to superhorizon scales: The onset of gravitational effects}
\label{sec:onset}

We now discuss the beta function for the effective potential in various regimes and compare it to its flat space (Minkowski) counterpart in order to pinpoint the specific effects of the space-time curvature.

\subsection{Minkowski regime}

The first case of interest is the regime of subhorizon scales $\kappa\gg1$, where all fluctuating modes are effectively heavy in units of the space-time curvature. One thus expects to recover the Minkowski limit of the flow equation. Indeed, using the asymptotic behavior $H_\nu(\kappa)\sim\sqrt{\frac{2}{\pi\kappa}}\exp\{i\kappa-i\frac{\pi}{2}(\nu+1/2)\}$ of the Hankel functions in \Eqn{nu_inv}, one finds $B_d(\nu,\kappa)\approx 8\kappa/\pi$. This leads to a beta function \eqn{pot_flow} identical to that obtained by deriving the flow equation directly in Minkowski space in the limit $\kappa^2\gg V_\kappa''$, as shown in Appendix~\ref{appsec:Mink}. 

\begin{figure}[t]
\begin{center}
\includegraphics[width=0.45\textwidth,]{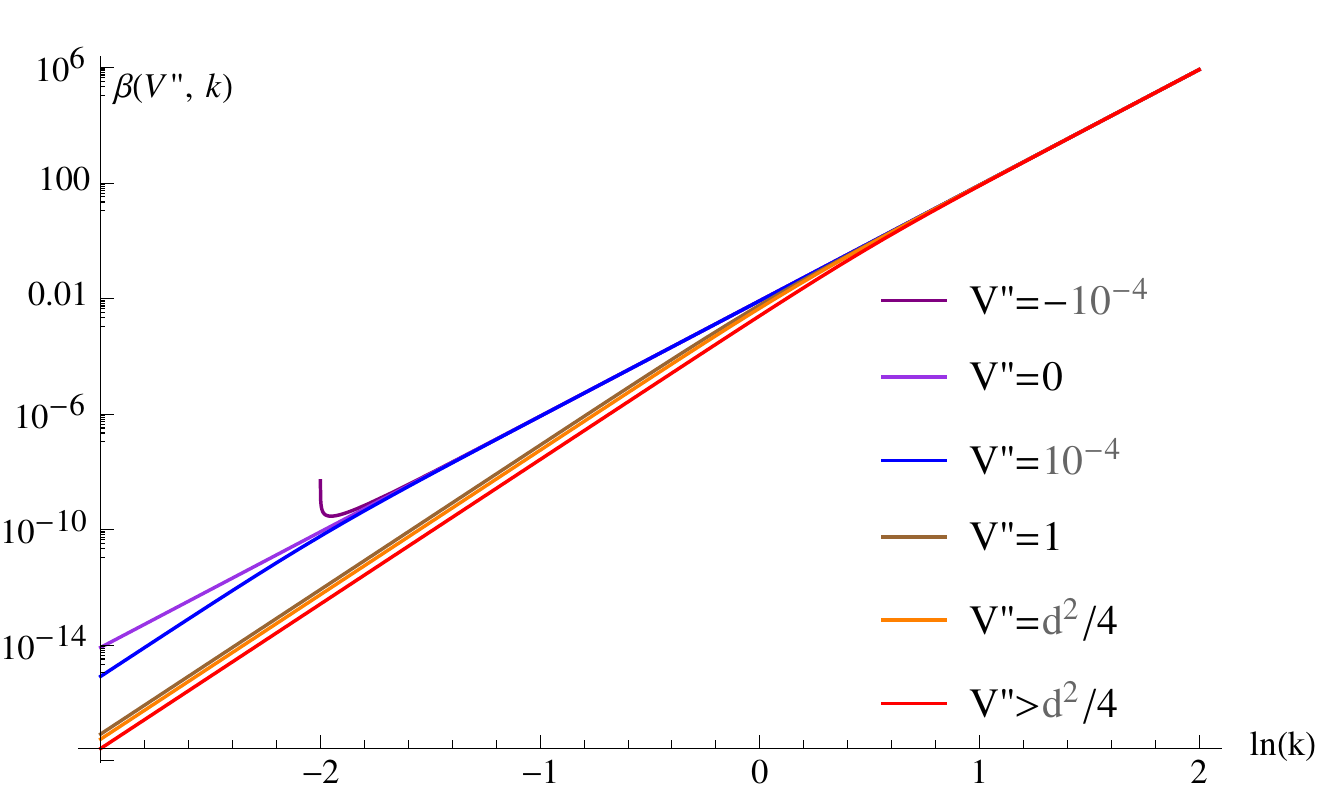}
\vspace{.3cm}
\includegraphics[width=0.45\textwidth,]{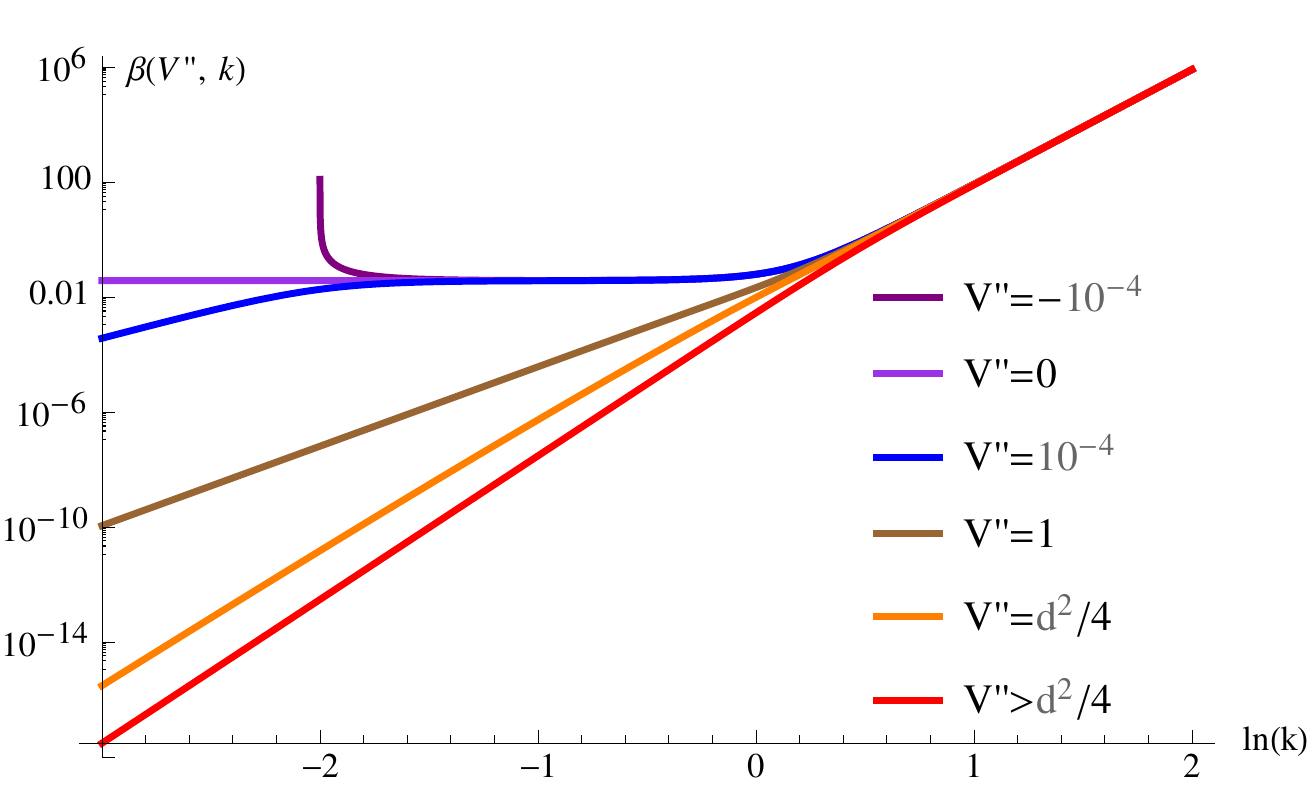}
\caption{The beta function $\beta(V'',\kappa)$ of the effective potential as a function of $\ln\kappa$ for different values of the potential curvature $V''$ in Minkowski (top) and de Sitter (bottom) space-times in $D=3+1$. The de Sitter beta function coincides with the Minkowski one for all values of $V''$ in the regime of subhorizon scales $\kappa\gg1$ and for all values of $\kappa$ when $V''\gg1$. Curvature effects become sizable on superhorizon scales for $V''\sim d^2/4$ [see \Eqn{eq:nu}]  and the de Sitter beta function is qualitatively different from the Minkowski one for small curvatures $V''\ll d^2/4$. In particular, its slope is dramatically reduced and even turns to zero for $V''\ll\kappa^2\ll1$ as a result of the gravitationally induced amplification of infrared fluctuations. This corresponds to the phenomenon of effective dimensional reduction described in the text. Also shown is the case of negative potential curvature, for which the beta function diverges as $\kappa^2\to V''$. In such regions of field space, the potential undergoes a strong RG flow which lowers the absolute value of the negative curvature.}
\label{beta_function}
\end{center}
\end{figure}

Similarly, for field values where the curvature of the potential $V_\kappa''\gg1$, one expects space-time curvature effects to be negligible for all $\kappa$.  In this case, the index $\nu_\kappa^2\approx -V_\kappa''$ and we obtain, using the properties of the Hankel function for imaginary index, $B_d(\nu,\kappa)\approx 8\sqrt{\kappa^2+V_\kappa''}/\pi$. The beta function \eqn{pot_flow} thus reads
\beq
\label{eq:betaMink}
 \beta\left(V_\kappa'' ,\kappa\right)\approx \frac{8C_d}{\pi}\frac{\kappa^{d+2}}{\sqrt{\kappa^2+V_\kappa''}},
\eeq
which is identical to the Minkowski beta function; see Appendix \ref{appsec:Mink}. The right-hand side of  \eqn{eq:betaMink} is plotted as a function of the RG scale $\kappa$ for various values of $V_\kappa''$ in the top panel of \Fig{beta_function}. 

\subsection{Infrared regime and dimensional reduction}

\begin{figure}[t]
\begin{center}
\includegraphics[width=0.45\textwidth,]{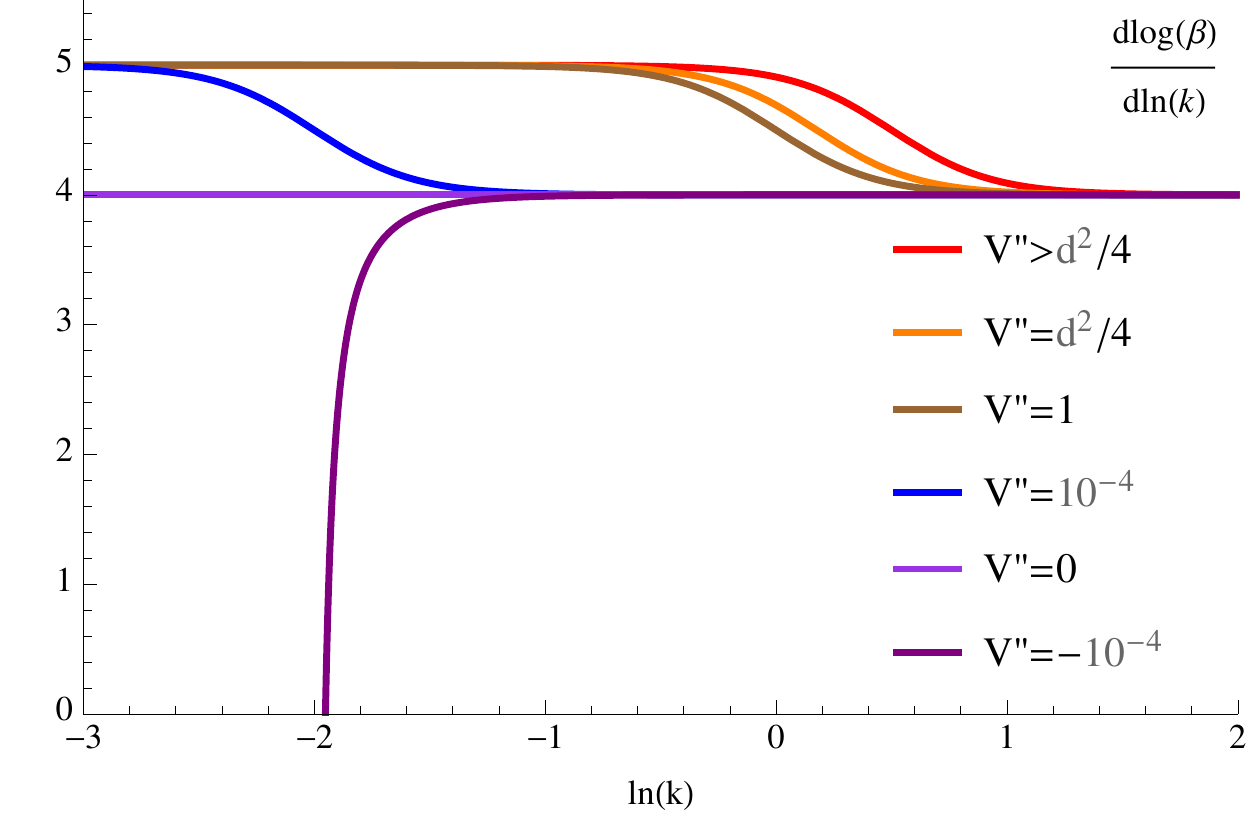}\\
\vspace{.3cm}
\includegraphics[width=0.45\textwidth,]{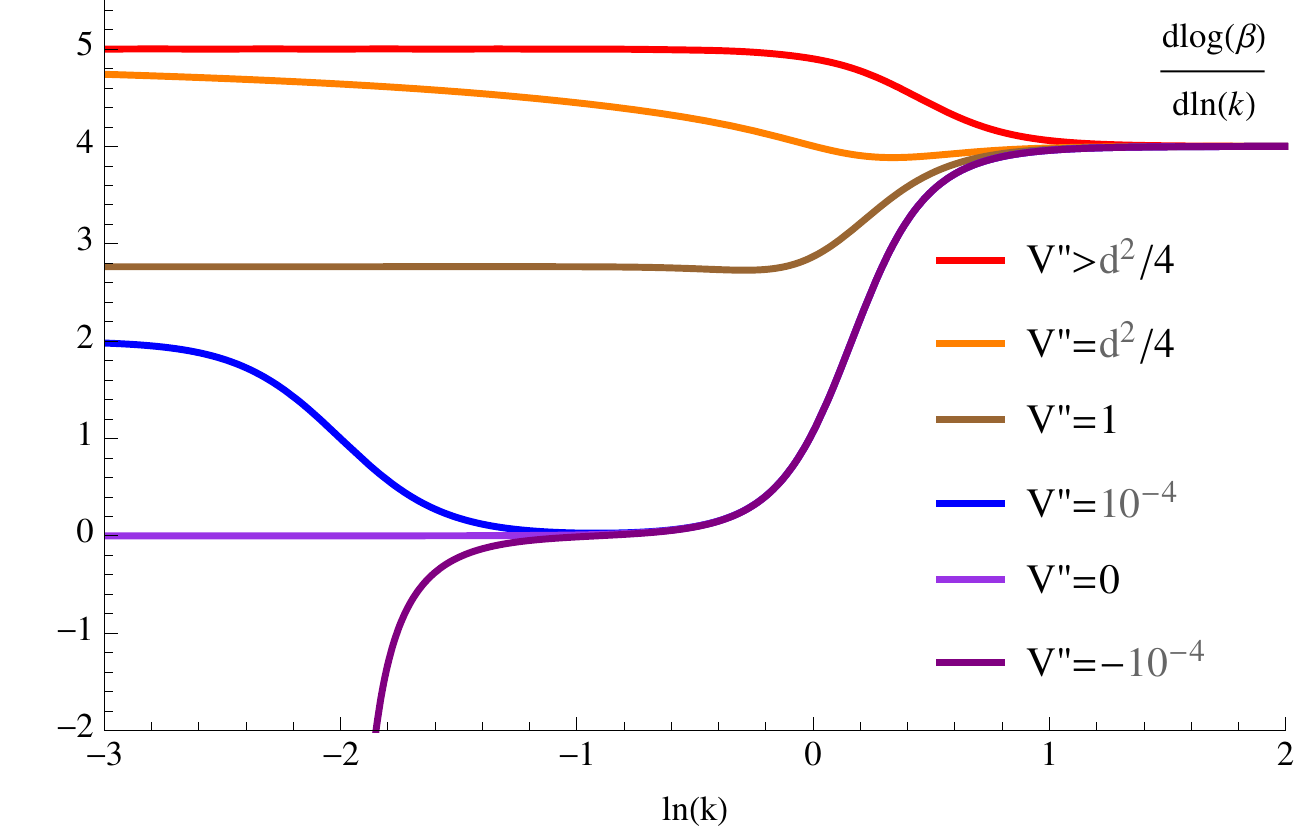}
\caption{The same as \Fig{beta_function} but for $\kappa\partial_\kappa\ln\beta(V'',\kappa)$. This shows the various power law behaviors in the different regimes of interest for the Minkowski (top) and the de Sitter (bottom) beta functions. In the former case, one has $\beta\sim\kappa^D$ for $\kappa^2\gg V''$ and $\beta\sim\kappa^{D+1}$ for $\kappa^2\ll V''$. In the de Sitter case, there is an extra dimensionful parameter and the structure is more complex. The Minkowski scaling is reproduced either for $\kappa^2\gg1$ or for $V''\gg1$ but there are strong modifications in the infrared regime $\kappa\ll1$ for $V''\lesssim d^2/4$. The gravitationally induced logarithmic and power law enhancements \eqn{eq:logenhancement} and \eqn{eq:powerenhancement} are clearly visible. The modified power law behavior in the infrared as compared to the flat space-time case results in an effective dimensional reduction up to the zero-dimensional scaling for $V''_\kappa\ll1$.}
\label{beta_slope}
\end{center}
\end{figure}

The Minkowski beta function \eqn{eq:betaMink} receives sizable corrections at superhorizon scales $\kappa\lesssim1$ when the curvature of the potential $V_\kappa''\lesssim d^2/4$. This corresponds to $\nu_\kappa^2$ increasing from (large) negative to positive values. For instance, for $V_\kappa''=d^2/4$ ($\nu_\kappa=0$), one has 
\begin{equation}
\label{eq:logenhancement}
B_d(0,\kappa) = \frac{4d^2}{\pi^2}\ln^2\left (\frac{\kappa}{2} \right ) - \frac{8d}{\pi^2}\ln\left(\frac{\kappa}{2}\right) + O(\kappa^0).
\end{equation}
This shows a (double) logarithmic enhancement as compared to the Minkowski case in the corresponding regime. This effect gets more pronounced as $V_\kappa''$ is further decreased ($\nu_\kappa$ is further increased to positive values). For $\nu\in\mathds{R}^+$ and $\kappa\lesssim1$, the Hankel functions $H_\nu(\kappa)\sim\frac{\Gamma(\nu)}{i\pi}(2/\kappa)^\nu$ and we obtain
\begin{equation}
\label{eq:powerenhancement}
B_d(\nu,\kappa) \approx d(d+2\nu)\frac{ \Gamma^2(\nu)}{\pi^2}  \left (\frac{2}{\kappa}\right)^{2\nu}\left[1+{\cal O}(\kappa^2)\right]
\end{equation}
The logarithmic enhancement of \Eqn{eq:logenhancement} is turned into a power law $\kappa^{-2\nu}$, which reflects the strong gravitational amplification of infrared fluctuations.
In the case of small potential curvature $|V_\kappa''|\ll1$, one has $\nu_\kappa\approx d/2$, and the beta function reads
\beq
\label{eq:IRbetafunc}
 \beta\left(V_\kappa'',\kappa\right)\approx \frac{1}{\Omega_{D+1}}\frac{ \kappa^{2} }{\kappa^2 + V_\kappa''},
\eeq
where we used $\Omega_{D+1}=4\pi^{d/2+1}/[d\Gamma(d/2)]$. The various regimes of the beta function in de Sitter space are illustrated in \Fig{beta_function} together with their Minkowski counterparts. 

Equation~\eqn{eq:IRbetafunc} reproduces the result of Ref.~\cite{Serreau:2013eoa} obtained directly in the infrared limit. As pointed out there, the beta function \eqn{eq:IRbetafunc} describes an effective Euclidean RG flow in zero space-time dimension.\footnote{A similar dimensional reduction phenomenon has been observed for fermionic degrees of freedom in spaces with constant negative curvature \cite{Gorbar:1999wa,Gies:2013dca}.} For instance, in the regime $V_\kappa''\ll\kappa^2\ll1$, the flow function $\beta(V_\kappa'',\kappa)\sim \kappa^0$, to be compared to the canonical scaling in $D$ dimensions $\sim\kappa^D$. Below we shall make this statement more precise by showing that the beta function \eqn{eq:IRbetafunc} describes a RG flow on the $D$-dimensional sphere $S_D$, that is, the Euclidean de Sitter space. As a measure of the effective dimensional reduction we show the logarithmic slope of the beta function in the various regimes of interest in \Fig{beta_slope}.

This effective dimensional reduction signals the fact that the solution of the flow equation governed by the beta function \eqn{eq:IRbetafunc} can be written as an effective zero-dimensional field theory. We introduce the following ordinary integral 
\begin{equation}
\label{eq:ordinaryint}
e^{-\Omega_{D+1}{\cal W}_\kappa(J)} {=} {\int} d^N \varphi \,e^{ -\Omega_{D+1}\left[V_{\rm eff}(\varphi)  + J_a\varphi_a +\frac{\kappa^2}{2}\varphi_a\varphi_a\right]},
\end{equation}
where $V_{\rm eff}(\varphi)$ is a function to be specified below. Repeating the steps leading to the flow equation \eqn{flot}, it is easy to check that the Legendre transform 
\beq
\label{eq:legendre}
 V_\kappa(\phi)={\cal W}_\kappa(J)-J_a\phi_a-\frac{\kappa^2}{2}\phi_a\phi_a,
\eeq
with $\partial{\cal W}_\kappa(J)/\partial J_a=\phi_a$, satisfies the flow equation \eqn{eq:IRbetafunc}. One can adjust the function $V_{\rm eff}(\varphi)$ so as to produce the appropriate initial conditions\footnote{In the case $N=1$, one can show that $V_{\rm eff}(\varphi)\approx V_{\kappa_0}(\varphi)$ if $V_{\rm eff}''(\varphi)\ll\kappa_0^2$. For arbitrary $N$, the inequality should be satisfied by the largest eigenvalue of the curvature matrix $\partial^2V_{\rm eff}(\varphi)/\partial\varphi_a\partial\varphi_b$.} for the infrared flow at a scale $\kappa_0\sim 1$. All solutions of the flow equation in the deep de Sitter regime can thus be written as \Eqn{eq:ordinaryint}. In particular, it is remarkable that, in this regime, the original $D$-dimensional Lorentzian theory, with complex weight $\exp(iS)$ eventually flows to a zero-dimensional Euclidean-like integral, with real weight $\exp(-\Omega_{D+1}V_{\rm eff})$.

\subsection{Relation to the stochastic approach}

The phenomenon of dimensional reduction described above is deeply related with the stochastic approach proposed by Starobinsky and Yokoyama in Ref.~\cite{Starobinsky:1994bd}. The latter is based on exploiting the specific aspects of the de Sitter kinematics to write down an effective theory for light fields on superhorizon scales. First, the large amplitude of quantum fluctuations on superhorizon scales implies that these behave as classical stochastic variables. Second, such fluctuations, of spatial size larger than the causal horizon are almost frozen in time and can essentially be described by a single degree of freedom\footnote{This can be generalized to take into account the field derivative $\partial_t\varphi_a(t)$ as an independent degree of freedom; see Ref.~\cite{Rigopoulos:2013exa}.}  $\varphi_a(t)$ in each direction in field space, with $t$ the cosmological time. Finally, because of the stationary gravitational redshift, this stochastic variable is sourced by the short wavelength (subhorizon) modes. The effective dynamics of the long wavelength modes is then described by an effective Langevin equation with delta-correlated noise~\cite{Starobinsky:1994bd,Beneke:2012kn}
\beq\label{eq:Langevin}
 \partial_t\varphi_a(t)+\frac{1}{d}\frac{\partial V_{\rm soft}(\varphi)}{\partial\varphi_a(t)}=\xi_a(t),
\eeq
where $V_{\rm soft}(\varphi)$ is the potential seen by the long wavelength modes (see below). 
Treating the short wavelength modes as noninteracting fields in the Bunch-Davies vacuum, one has, generalizing the calculation of \cite{Starobinsky:1994bd,Beneke:2012kn} to arbitrary $N$,
\beq
 \langle\xi_a(t)\xi_b(t')\rangle=\frac{\Gamma(d/2)}{2\pi^{{d\over2}+1}}\delta_{ab}\delta(t-t').
\eeq
Using standard manipulations, \Eqn{eq:Langevin} can be turned into the following Focker-Planck equation for the probability distribution ${\cal P}(\varphi,t)$ of the stochastic process
\beq
\label{eq:FP}
 \partial_t{\cal P}=\frac{1}{d}\frac{\partial}{\partial\varphi_a}\left\{\frac{\partial V_{\rm soft}}{\partial\varphi_a}{\cal P}+\frac{1}{\Omega_{D+1}}\frac{\partial{\cal P}}{\partial\varphi_a}\right\}.
\eeq
The latter admits an $O(N)$-symmetric stationary attractor solution at late times (i.e., in the deep infrared), given by
\beq
 {\cal P}(\varphi)\propto\exp\big\{-\Omega_{D+1}V_{\rm soft}(\varphi)\big\}.
\eeq
Equal-time correlation functions on superhorizon scales can then be computed as moments of this distribution. This coincides with the outcome \eqn{eq:ordinaryint} of the above RG analysis in the limit $\kappa\to0$ provided one identifies $V_{\rm soft}(\varphi)=V_{\rm eff}(\varphi)\approx V_{\kappa_0}(\varphi)$. For instance, one has
\beq
 \langle\varphi_a\varphi_b\rangle=\frac{\int d^N\varphi\,\varphi_a\varphi_b\,{\cal P}(\varphi)}{\int d^N\varphi\,{\cal P}(\varphi)}=\left.\frac{1}{\Omega_{D+1}}\frac{\partial^2{\cal W}_{\kappa=0}(J)}{\partial J_a\partial J_b}\right|_{J=0}.
\eeq
The relevant potential to be used in the stochastic approach is thus not the microscopic one (at the UV scale $\Lambda$) but the one evolved down to the horizon scale $\kappa_0$, which makes perfect physical sense.

The present NPRG approach thus sheds a new light on the basic principles underlying the stochastic approach. Moreover, it clarifies the relation between the stochastic approach and the Euclidean de Sitter approach, as we now discuss.

\subsection{Relation to Euclidean de Sitter space}

Another interesting consequence of the dimensional reduction concerns the relation between Lorentzian and Euclidean de Sitter spaces, the latter being nothing but the $D$-dimensional sphere $S_D$. It has been pointed out in \cite{Beneke:2012kn} that, for what concerns the calculation of static quantities (e.g., equal-time correlators) on superhorizon scales, the nonperturbative physics of the zero mode on the sphere reproduces the results of the stochastic approach. However, the origin of this result has remained unclear. 

The present NPRG approach allows us to clarify this point. As we have discussed above, the stochastic approach emerges as the result of the effective dimensional reduction of the RG flow due to strong enhancement of infrared fluctuations in the Lorentzian case. A similar dimensional reduction takes place in the Euclidean case for more obvious reasons since the sphere is compact.\footnote{Dimensional reduction is spaces with compact dimension has been studied in \cite{Hu:1986cv}. The number of effective dimension is simply given by the number of noncompact dimensions.} The spectrum of the theory is thus discrete and all heavy modes decouple for scales below the first excited level, leaving the zero mode as the only fluctuating degree of freedom. 

The effective dimensional reduction for a scalar field theory ($N=1$) on the sphere has been studied in detail by means of NPRG techniques in Ref.~\cite{Benedetti:2014gja}. There the author finds, employing the LPA and a Litim regulator, that the beta function for the effective potential on length scales larger than the sphere radius exactly reproduces the one obtained in \cite{Serreau:2013eoa} for the Lorentzian theory on superhorizon scales, \Eqn{eq:IRbetafunc}. Below, we provide a short alternative description of the origin of the dimensional reduction on the sphere.

The generating functional for connected correlation functions is given by
\beq 
\label{eq:sphere}
 e^{-\bar W_\kappa[J]}=\!\int\!{\cal D}\varphi\exp\left(-\bar S[\varphi]-\Delta \bar S_\kappa[\varphi]-\int_xJ_a\varphi_a\right),
\eeq
where we denote Euclidean quantities by an overall bar (we do not need to be more precise here) and $\int_x$ is the invariant integration on the unit sphere $S_D$. One decomposes the fields on the discrete basis of eigenfunctions of the corresponding Laplacian operator
\beq
 \varphi_a(x)=\sum_{\vec L}\varphi_{a,\vec L}Y_{\vec L}(x),
\eeq
where $\vec L=(L,L_{D-1},\ldots,L_1)$ is a vector of integer numbers with $L\ge L_{D-1}\ge\ldots\ge|L_1|$ and where the spherical harmonics satisfy
\beq
 \square_{S_D} Y_{\vec L}(x)=-\lambda_LY_{\vec L}(x),
\eeq
with $\lambda_L=L(L+D-1)$, and are normalized as
\beq
 \int_xY_{\vec L}^*(x)Y_{\vec L'}(x)=\delta_{\vec L,\vec L'}.
\eeq
The zero mode is the constant $Y_0=1/\sqrt{\Omega_{D+1}}$, with $\Omega_{D+1}$ the volume of the unit sphere $S_D$.
The infrared regulator in \Eqn{eq:sphere} can be written as
\beq
 \Delta \bar S_\kappa[\varphi]=\frac{1}{2}\sum_{a,\vec L} \bar R_\kappa(L)\,|\varphi_{a,\vec L}|^2
\eeq
where the function $\bar R_\kappa(L)$  provides a large effective mass for modes such that $\lambda_L\lesssim\kappa^2$. Because the spectrum is discrete, it is essentially constant for scales below the first nonzero mode $\kappa^2\lesssim D$. For a potential curvature lower than the first level, $V''\lesssim D$, and for scales $\kappa^2\lesssim D$, the nonzero modes effectively behave as heavy modes and decouple in the flow equation. The physics of the zero mode is nonperturbative and must be treated separately \cite{Rajaraman:2010xd,Beneke:2012kn}. 

For instance, employing the following regulator 
\beq
 \bar R_\kappa(L)=\Big(\kappa^2-\lambda_L\Big)\,\theta\Big(\kappa^2-\lambda_L\Big),
\eeq
one has $\bar R_\kappa(L)=\kappa^2\delta_{L,0}$ for $\kappa^2<D$. Writing the field as
\beq
 \varphi_a(x)=\bar\varphi_a+\hat\varphi_a(x),
\eeq
with $\bar\varphi_a=\varphi_{a,0}Y_0=\int_x\varphi_a(x)/\Omega_{D+1}$, we define the generating function for the fluctuations of the zero mode as $\overline W_\kappa[J={\rm const.}]=\Omega_{D+1}\overline{\cal W}_\kappa(J)$, which reads
\beq
\label{eq:sphere2}
 e^{-\Omega_{D+1}\overline{\cal W}_\kappa(J)}=\int d^N\bar\varphi\,e^{-\Omega_{D+1}\left[\bar V_{\rm eff}(\bar\varphi)+{\kappa^2\over2}\bar\varphi_a\bar\varphi_a+J_a\bar\varphi_a\right]}.
\eeq
Here we wrote ${\cal D}\varphi=d^N\bar\varphi{\cal D}\hat\varphi$ and we defined the effective potential for the zero mode as
\beq
  e^{-\Omega_{D+1}\bar V_{\rm eff}(\bar\varphi)}=\int{\cal D}\hat\varphi\,e^{-\bar S[\varphi]}.
\eeq 
Equation \eqn{eq:sphere2} coincides with the Lorentzian result \Eqn{eq:ordinaryint}---and thus with the stochastic approach as discussed above---provided one identifies the respective effective potentials $V_{\rm eff}$ and $\bar V_{\rm eff}$.

\section{Large-$N$ limit}
\label{sec:largeN}

We now discuss the actual RG flow from subhorizon to superhorizon scales. We first consider the limit of a large number of field components, $N\to\infty$, for which the flow equation for the potential is exactly given by the LPA \cite{D'Attanasio:1997he} and can be solved analytically in the interesting infrared regime. Furthermore, as we shall see later, the large-$N$ limit correctly captures the qualitative behavior of the finite $N$ case. 

For $N\to\infty$, only the transverse modes contribute to the flow equation \eqn{eq:flot_N}, which becomes
\beq
\label{large_N_flow}
\dot U_\kappa(\rho) =  \beta\left(U'_\kappa(\rho),\kappa\right),
\eeq
with the beta function given by Eqs.~\eqn{pot_flow} and \eqn{nu_inv}. A standard trick \cite{Tetradis:1995br,Blaizot:2005xy} is to rewrite this equation in terms of the function $\rho_\kappa(W)$ defined by the relation\footnote{This assumes that the function $U_\kappa'(\rho)$ or, equivalently, $\rho_\kappa(W)$, is invertible. It is easy to check that $\dot\rho_\kappa(W)$ in \Eqn{eq:flowrhokappa} is a decreasing function of $W$: $\dot\rho_\kappa'(W)\le0$. Here, we shall consider cases where the initial condition at the scale $\kappa=\Lambda$ is a monotonous---thus invertible---function with $\rho_\Lambda'(W)\ge0\,\,\forall \,W$. It follows that $\rho_{\kappa\le\Lambda}'(W)\ge0\,\,\forall \,W$ and hence the function $\rho_\kappa(W)$ is invertible for all $\kappa\le\Lambda$.} $U_\kappa'\big(\rho_\kappa(W)\big)=W$. One thus has $\dot\rho_\kappa(W)=-\dot U_\kappa'(\rho)/U_\kappa''(\rho)|_{\rho=\rho_\kappa(W)}$ as well as $U_\kappa''\big(\rho_\kappa(W)\big)\rho_\kappa'(W)=1$ and the flow takes the following explicit expression
\beq
\label{eq:flowrhokappa}
 \dot\rho_\kappa(W)=-\partial_W\beta(W,\kappa).
\eeq
An important property of this flow equation is that, because the $\kappa$-dependence of the right-hand side is explicit, the coefficients of the Taylor expansion of $\rho_\kappa(W)$ in $W$, e.g., around $W=0$, all have independent RG flows.

A typical initial condition at the UV scale $\kappa=\Lambda$ is $U_\Lambda(\rho)=m_\Lambda^2\rho+\lambda_\Lambda\rho^2/2$, that is, $\rho_\kappa(W)=(W-m_\Lambda^2)/\lambda_\Lambda$. Here, the parameter $m_\Lambda^2$ can be of any sign and $\lambda_\Lambda\ge0$. The flow in the UV regime $\kappa\gtrsim1$ is described by the Minkowski beta function \eqn{eq:betaMink} and one gets
\beq
\label{eq:UVflowN}
 \rho_\kappa(W)=\rho_{\Lambda}(W)-\frac{4C_d}{\pi}\int_{\kappa}^{\Lambda}dk\frac{ k^{d+1}}{\big(k^2+W\big)^{3/2}}.
\eeq
For theories deep in the symmetric phase, where $U_\kappa'(\rho)\gg1\,\,\forall\rho\ge0$, the flow eventually freezes out in the Minkowski regime at a scale $\kappa^2\sim U_\kappa'(0)$. More interesting are the cases of theories either close to criticality or deep in the broken phase, for which there exists a significant region in field space where\footnote{This stems from the fact that, unlike the interpolating potential $U_\kappa(\rho)$, the regulated potential $U_\kappa(\rho)+R_\kappa(0)\rho$ is a convex function of $\varphi_a$ \cite{Berges:2000ew,Delamotte:2007pf}. Indeed, it is the Legendre transform of the generating functional \eqn{functional} for constant sources $W[J={\rm const.}]$, which is a convex function of $J_a$. Note that this assumes that the infrared regulator $R_\kappa(p)$ indeed completely regulates the theory at all scales $\kappa$. With the regulator \eqn{reg}, this implies that a possibly concave region is such that the negative curvature never exceeds the IR cutoff scale: $\kappa^2+U_\kappa'(\rho)>0$.} $|U_\kappa'(\rho)|\lesssim1$ down to scales $\kappa\sim1$. This is the case where we expect important gravitational effects. In the region $W\ll1$, the Minkowski flow \eqn{eq:UVflowN} reads
\beq
\label{eq:MinkflowUVlargeN}
 \rho_\kappa(W)=\frac{W-m_\kappa^2}{\lambda_\kappa}+{\cal O}\left(W^2\right),
\eeq 
where
\begin{align}
\label{eq:massflow}
 \frac{m_\kappa^2}{\lambda_\kappa}&=\frac{m_\Lambda^2}{\lambda_\Lambda}+\frac{4C_d}{\pi}\frac{\Lambda^{D-2}-\kappa^{D-2}}{D-2}\\
\label{eq:couplingflow}
 \frac{1}{\lambda_\kappa}&=\frac{1}{\lambda_\Lambda}+\frac{6C_d}{\pi}\frac{\Lambda^{D-4}-\kappa^{D-4}}{D-4}.
\end{align}

For infrared scales $\kappa\ll1$, the flow of the part of the potential where $|U_\kappa'(\rho)|\ll1$ is described by the dimensionally reduced beta function \eqn{eq:IRbetafunc} and one gets
\beq
\label{eq:hahaha}
 \rho_\kappa(W)=\rho_{\kappa_0}(W)+\frac{1}{2\Omega_{D+1}}\left\{\frac{1}{\kappa_0^2+W}-\frac{1}{\kappa^2+W}\right\},
\eeq
where $\kappa_0\sim1$ denotes the horizon scale. Using the approximate UV flow \eqn{eq:MinkflowUVlargeN} down to the scale $\kappa_0$, we have $U_{\kappa_0}(\rho)\approx m_{\kappa_0}^2\rho+\lambda_{\kappa_0}\rho^2/2$ and \Eqn{eq:hahaha} can be rewritten as
\beq
\label{eq:gggg}
 \left(U_\kappa'+\kappa_0^2\right) \left(U_\kappa'+\kappa^2\right)\left(U_\kappa'-U_{\kappa_0}'\right)=\frac{\lambda_{\kappa_0}}{2\Omega_{D+1}}\left(\kappa_0^2-\kappa^2\right).
\eeq
Under the above assumptions, we have $U_{\kappa}'(\rho)\ll\kappa_0^2$ in the relevant region of the potential and \Eqn{eq:gggg} becomes a second order polynomial equation for $U_\kappa'$. The latter can easily be solved and integrated in $\rho$. Introducing the function $\tilde U_\kappa(\rho)=U_\kappa(\rho)+\kappa^2\rho$, we obtain
\begin{align}
\label{eq:potflowlargeNIR1}
 \tilde U_\kappa(\rho)-\tilde U_\kappa(0)&=\frac{M^4_\kappa(\rho)-M^4_\kappa(0)}{2\lambda_{\kappa_0}}\nn&+\frac{1}{2\Omega_{D+1}}\left(1-\frac{\kappa^2}{\kappa_0^2}\right)\ln\frac{M^2_\kappa(\rho)}{M^2_\kappa(0)},
\end{align}
where the curvature term  $M^2_\kappa(\rho)=\tilde U_\kappa'(\rho)=U_\kappa'(\rho)+\kappa^2$ is given by\footnote{Notice that $\tilde U_\kappa(\rho)$ is nothing but the Legendre transform potential mentioned earlier. We check that the latter is a convex function of $\rho$ all along the (infrared) flow: $M_\kappa^2(\rho)=U_\kappa'(\rho)+\kappa^2>0$. Finally, we recall that the expressions \eqn{eq:potflowlargeNIR1} and \eqn{eq:potflowlargeNIR2} are valid provided $M^2_\kappa(\rho)\ll1$.}
\begin{align}
\label{eq:potflowlargeNIR2}
 M^2_\kappa(\rho)&=\frac{m_{\kappa_0}^2+\lambda_{\kappa_0}\rho+\kappa^2}{2}\nn&+\sqrt{\left(\frac{m_{\kappa_0}^2+\lambda_{\kappa_0}\rho+\kappa^2}{2}\right)^2+\frac{\lambda_{\kappa_0}}{2\Omega_{D+1}}\left(1-\frac{\kappa^2}{\kappa_0^2}\right)}.
\end{align}
For $\kappa=0$, this reproduces the result of Ref.~\cite{Serreau:2011fu}, obtained by a direct calculation of the effective potential in the limit $N\to\infty$. We mention that the above result for the running potential in the infrared regime can equivalently be obtained by a direct calculation of the integral \eqn{eq:ordinaryint} using standard large-$N$ techniques.

\begin{figure}[t]
\begin{center}
\includegraphics[width=0.4\textwidth,]{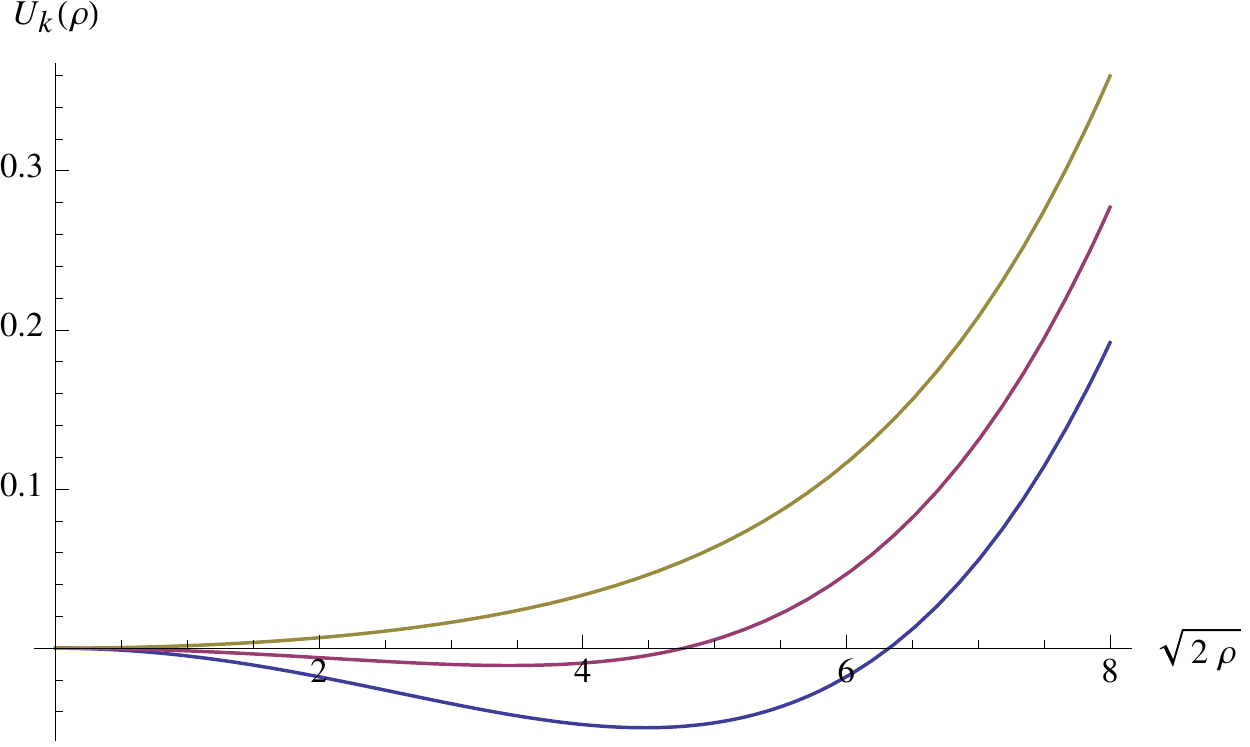}\\
\caption{The effective potential $U_\kappa(\rho)$ in the limit $N\to\infty$ [see \Eqn{eq:potflowlargeNIR1}] in $D=3+1$ as a function of the radial variable $\sqrt{2\rho}$ in field space for (from bottom to top) $\kappa=1,0.1,0$. The parameters at the horizon scale $\kappa_0=1$ are taken as $m_{\kappa_0}^2=-0.01$, and $\lambda_{\kappa_0}=0.001$.}
\label{flow_pot_IR_largeN}
\end{center}
\end{figure}

\subsection{Symmetry restoration}

Let us discuss some consequences of the findings of the previous sections. As pointed out in Ref.~\cite{Serreau:2013eoa}, an important consequence of the effective dimensional reduction of the RG flow in the infrared regime is the fact that any spontaneously broken symmetry gets radiatively restored. This is easily understood from the fact that the generating function of the effective zero-dimensional field theory given by the ordinary integral \Eqn{eq:ordinaryint} is analytic and cannot present a spontaneously broken phase. In the limit $N\to\infty$, this phenomenon of symmetry restoration along the flow in the infrared regime can be seen on the exact solution, Eqs.~\eqn{eq:potflowlargeNIR1} and \eqn{eq:potflowlargeNIR2}, as illustrated on \Fig{flow_pot_IR_largeN}.

\begin{figure}[t]
\begin{center}
\includegraphics[width=0.45\textwidth, trim = 20 8 10 10,clip]{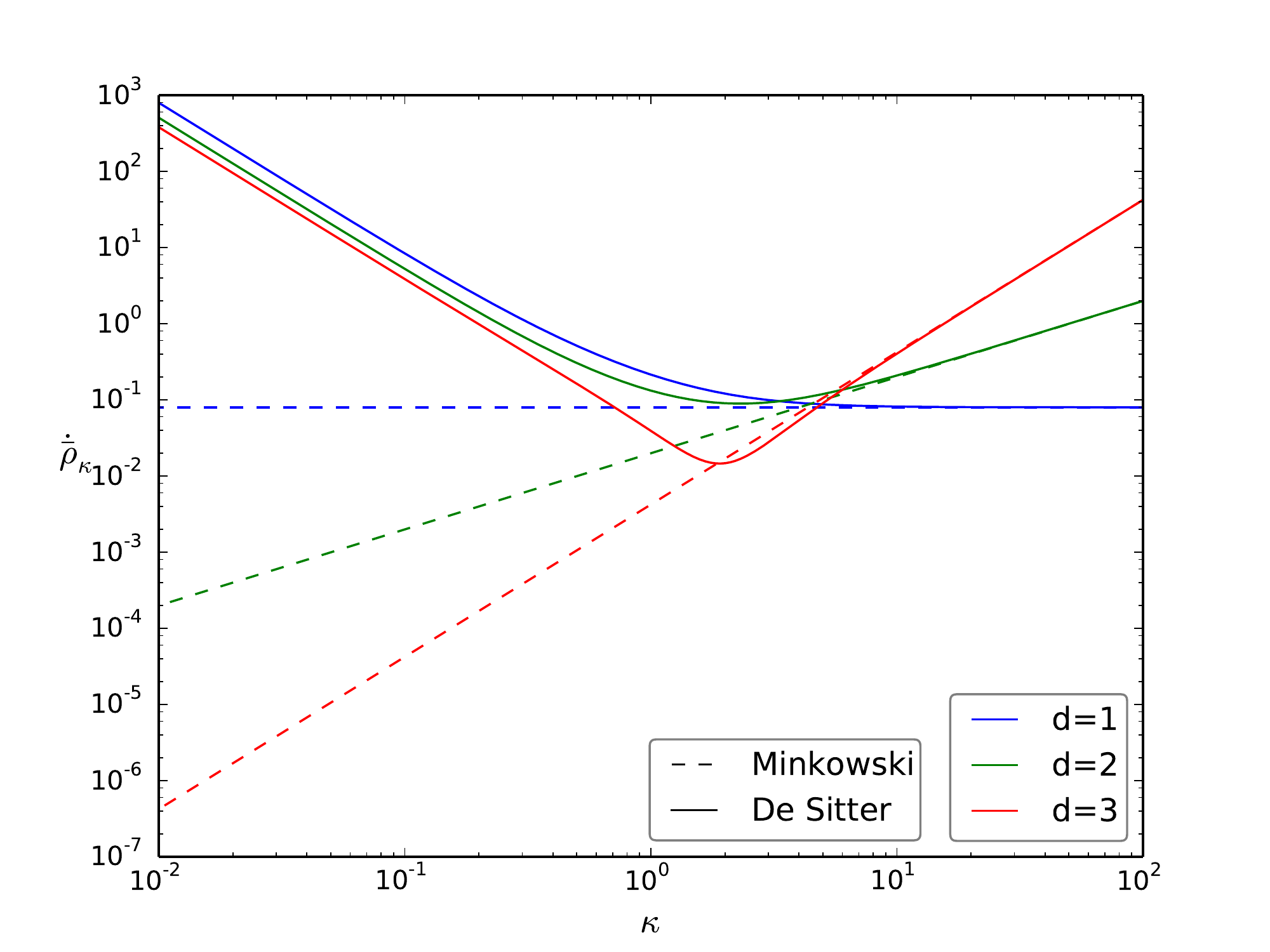}
\caption{The beta functions for the minimum of the potential in the large-$N$ limit in de Sitter (plain lines) and Minkowski (dashed lines) space-times in $D=d+1$ dimensions. The de Sitter and Minkowski beta functions coincide in the regime of subhorizon scales $\kappa\gg1$, where they behave as a power law $\kappa^{D-2}$. Significant deviations occur for scales close to the horizon, $\kappa\sim1$. As a result of the strong gravitational enhancement of infrared fluctuations, the de Sitter beta functions switch to a common $\kappa^{-2}$ behavior for superhorizon scales, which signals an effective zero-dimensional flow.}
\label{largeNflow}
\end{center}
\end{figure}

The analysis of Ref.~\cite{Serreau:2013eoa} was restricted to the deep infrared regime, where the flow is already dimensionally reduced. Here, we extend this discussion and we consider the complete flow from subhorizon to superhorizon scales. This allows us to study how a possible broken phase in the Minkowski regime gets restored once gravitational effects become important in the infrared regime. We follow the flow of the minimum $\bar\rho_\kappa$ of the potential, defined as \mbox{$U_\kappa'(\bar\rho_\kappa)=0$} or, equivalently, as $\bar\rho_\kappa=\rho_\kappa(W=0)$. As explained above, the RG flow of $\bar\rho_\kappa$ is independent of that of other couplings. We have, from \Eqn{eq:flowrhokappa},
\beq
\label{eq:flowzero}
 \dot{\bar\rho}_\kappa=-\left.\partial_W\beta(W,\kappa)\right|_{W=0}.
\eeq
The right-hand side can be evaluated in closed form for each dimension $d$. For instance, we get 
\begin{align}
\label{eq:rhobar1}
\dot {\bar \rho}_\kappa &\underset{d=1}{=}  \frac{1 }{4 \pi}  \left \{ \frac{1}{\kappa^2}+3  +2g(2\kappa) - 4\kappa f(2\kappa) \right \} \\
\label{eq:rhobar2}
\dot {\bar \rho}_\kappa &\underset{d=2}{=}  \frac{1 }{32}  \bigg \{  \left(4+2 \kappa^2  \right)\left|H_1(\kappa)\right|^2 -\kappa^2 \left|H_0(\kappa)\right|^2  \bigg \} \\
\label{eq:rhobar3}
\dot {\bar \rho}_\kappa &\underset{d=3}{=} \frac{1 }{72 \pi^2}  \Bigg\{  \frac{27}{\kappa^2} + 15 - \kappa^2 + 2\!\left(9-16\kappa^2 \right)\!g(2\kappa) \nn
&  \,\, -4\kappa\! \left(9 -2\kappa^2 \right)\!f(2\kappa)  \Bigg\} ,
\end{align}
where the real functions $g$ and $f$ are defined as 
\beq
g(x) + if(x) = \int_0^\infty du\frac{e^{iu}}{u+x} \quad{\rm for}\quad x>0.
\eeq
The functions \eqn{eq:rhobar1}--\eqn{eq:rhobar3} are plotted in \Fig{largeNflow} along with their equivalents in Minkowski space. As before, the subhorizon regime is governed by the Minkowski beta function \eqn{eq:betaMink}, which yields
\beq
\label{eq:Minkflowmin}
 \dot{\bar \rho}_\kappa\approx\frac{4 C_d}{\pi}\kappa^{D-2}\quad{\rm for}\quad\kappa\gg1.
\eeq
One easily checks that the functions \eqn{eq:rhobar1}--\eqn{eq:rhobar3} are indeed given by the above formula in this regime. One sees in \Fig{largeNflow} that gravitational corrections become significant for $\kappa\sim1$ and dramatically modify the flow for $\kappa\ll1$, where the functions  \eqn{eq:rhobar1}--\eqn{eq:rhobar3} acquire the same slope in all dimensions. This signals the effective dimensional reduction discussed above. Indeed, inserting the beta function \eqn{eq:IRbetafunc} in \Eqn{eq:flowzero}, we obtain
\beq
\label{eq:IRflowmin}
\dot{\bar \rho}_\kappa  \approx\frac{1}{\Omega_{D+1} \kappa^2}\quad{\rm for}\quad\kappa\ll1,
\eeq
which reproduces the small $\kappa$ behavior of Eqs.~\eqn{eq:rhobar1}--\eqn{eq:rhobar3}.

In the Minkowski regime, the flow \eqn{eq:Minkflowmin} integrates to
\beq
\label{eq:Minkflowminres}
 \bar\rho_\kappa=\bar\rho_{\Lambda}-\frac{4C_d}{\pi}\frac{\Lambda^{D-2}-\kappa^{D-2}}{D-2}\quad{\rm for}\quad1\lesssim\kappa\le\Lambda
\eeq
and we recover the following known facts. First, in $D=2$, the minimum of the potential would reach zero at a finite scale $\kappa=\Lambda\exp(-4\pi\bar\rho_\Lambda)$ for any initial condition and the Minkowski theory has no phase of spontaneously broken symmetry. In contrast, in $D>2$, the Minkowski theory reaches a phase of broken symmetry in the limit $\kappa\to0$ if $\bar\rho_\Lambda>\rho_c=4C_d\Lambda^{D-2}/[\pi(D-2)]$. For $\bar\rho_\Lambda=\rho_c$, the Minkowski theory is critical.

\begin{figure}
\begin{center}
\includegraphics[width=0.45\textwidth,trim = 20 8 10 10, clip]{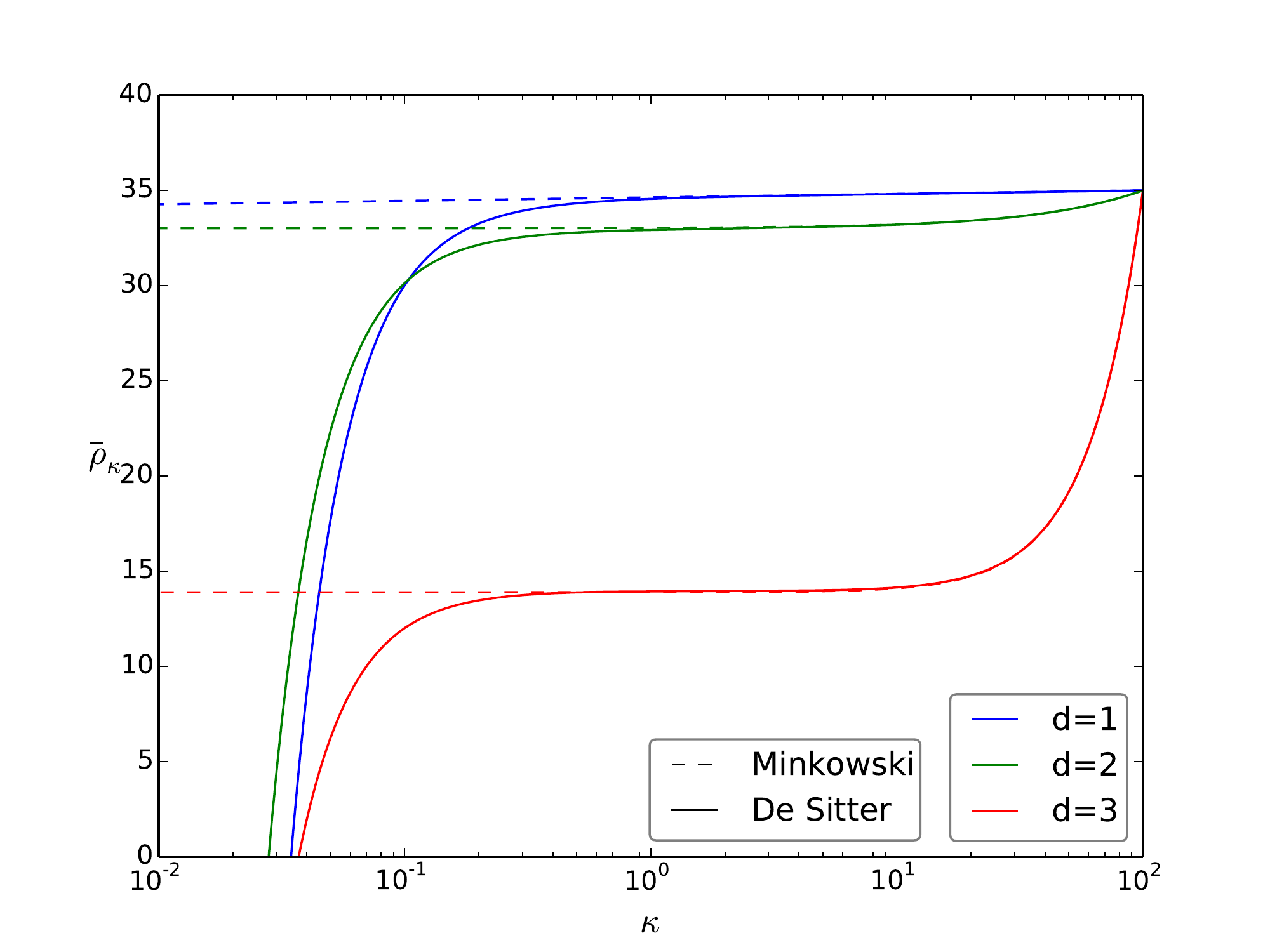}
\caption{The flow of the minimum of the potential in de Sitter (plain lines) and Minkowski (dashed lines) space-times obtained by a direct integration of the beta functions shown in \Fig{largeNflow}. The initial condition $\bar \rho_\Lambda$ at the scale $\Lambda=10^2$ is chosen such that the Minkowski theories in $D>2$ are in the broken phase. We clearly see the effects of gravitationally amplified infrared modes in de Sitter space which quickly restore the symmetry as soon as $\kappa\lesssim1$. In the case $D=2$, the Minkowski flow slowly restores the symmetry with a logarithmic flow. Infrared de Sitter effects lead to a much faster (power law) symmetry restoration.}
\label{largeNflow2}
\end{center}
\end{figure}

These matters are drastically changed in de Sitter space for $\kappa\lesssim1$. In that regime, the flow \eqn{eq:IRflowmin} integrates to
\beq
\label{eq:IRflowminres}
  \bar\rho_\kappa=\bar\rho_{\kappa_0}+\frac{1}{2\Omega_{D+1}}\left(\frac{1}{\kappa_0^2}-\frac{1}{\kappa^2}\right)\quad{\rm for}\quad\kappa\le\kappa_0\lesssim1
\eeq
and one sees that the minimum of the potential reaches zero at a finite scale so the theory always ends up in the symmetric phase at $\kappa=0$. The flow of the minimum of the potential is shown in \Fig{largeNflow2} in various dimensions for an initial condition which would result in a broken phase in Minkowski space in both $D=3$ and $D=4$. The plain curves are obtained by integrating the complete flow equations \eqn{eq:rhobar1}--\eqn{eq:rhobar3} and are compared to the corresponding flow in Minkowski space. We see that, even in the case $D=2$, where the Minkowski flow would eventually reaches the symmetric phase, gravitational effects make a qualitative difference and dramatically speed up symmetry restoration. Finally, we mention that the result of the numerical integration of Eqs.~\eqn{eq:rhobar1}--\eqn{eq:rhobar3} in that case is quantitatively well described by Eqs.~\eqn{eq:Minkflowminres} and \eqn{eq:IRflowminres} with a matching point at $\kappa_0=1$.

\subsection{Mass (re)generation}

As we have seen previously, a theory with a large mass gap in units of the space-time curvature does not feel any de Sitter effects and is essentially described by the Minkowski flow all the way to the deep infrared. Space-time curvature plays a nontrivial role when there are light excitations $m_{\kappa_0}\lesssim\kappa_0$ at the horizon scale $\kappa_0\sim1$. This is the case for theories which are nearly critical ($\bar\rho_\Lambda\approx\rho_c$) or in the broken phase ($\bar\rho_\Lambda\gtrsim\rho_c$) at subhorizon scales.

We thus consider initial conditions at the UV scale $\Lambda$ such that $\bar\rho_\Lambda\ge\rho_c$. The flow of the minimum of the potential has been described in the previous subsection. As long as it is nonzero, the mass of the transverse Goldstone modes vanish identically $m_{t,\kappa}^2=U_\kappa'(\bar\rho_\kappa)=0$ whereas the mass of the longitudinal mode is given by $m_{l,\kappa}^2=2\lambda_\kappa\bar\rho_\kappa$, where $\lambda_\kappa=U_\kappa''(\bar\rho_\kappa)$. Once the symmetry gets restored, the minimum of the potential stays at $\bar\rho_\kappa=0$ and the transverse and longitudinal masses become degenerate: $m_{t,\kappa}^2=m_{l,\kappa}^2=U_\kappa'(0)\equiv m_\kappa^2$.

\begin{figure}[t]
\begin{center}
\includegraphics[width=0.4\textwidth]{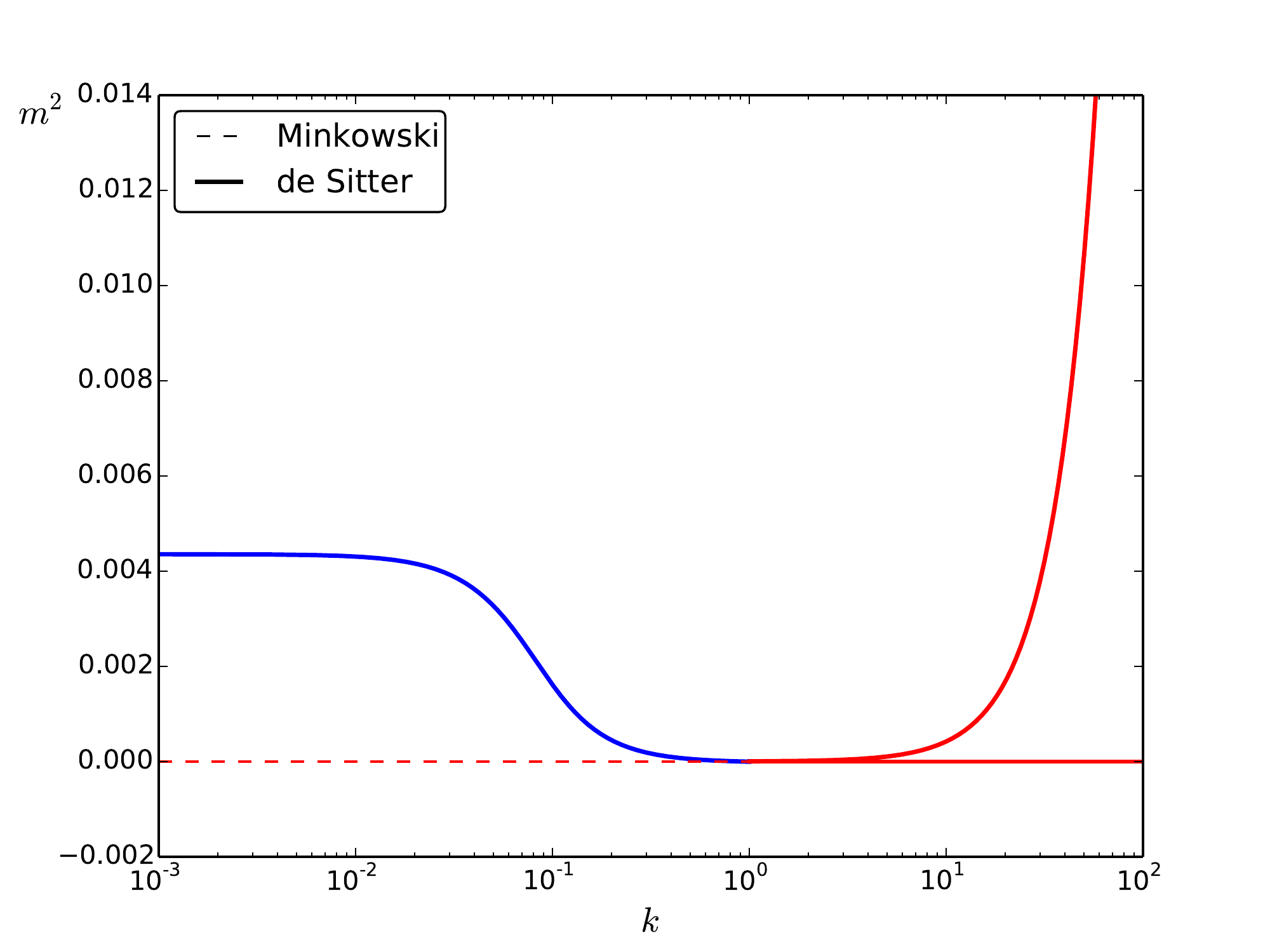}\\
\vspace{.3cm}
\includegraphics[width=0.4\textwidth]{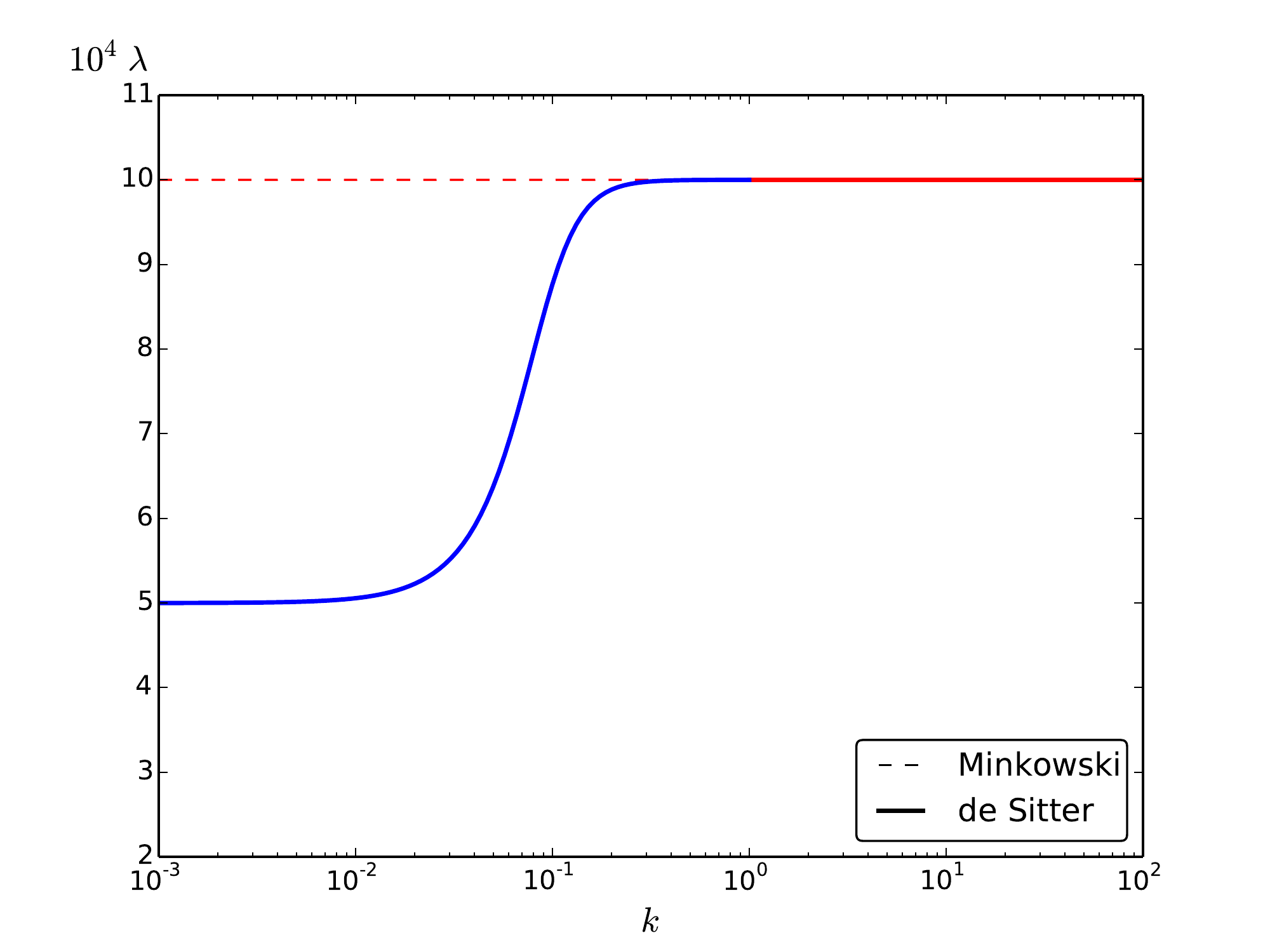}
\caption{Flow of the (would-be) critical theory in $D=3+1$ in the large-$N$ limit (see text). The initial conditions at the scale $\Lambda=10^2$ are $\rho_\Lambda=\rho_c=625/(3\pi^2)\approx21$ and $\lambda_\Lambda=10^{-3}$. The upper panel shows the flow of longitudinal and transverse masses. The transverse Goldstone mass is zero and the longitudinal one decreases until symmetry restoration at $\kappa=\kappa_0\sim1$. For lower scales both masses agree and a nontrivial infrared gap is generated (blue curve). The lower panel shows the flow of the coupling constant $\lambda_\kappa=U_\kappa''(\bar\rho_\kappa)$. The UV flow is very slow (logarithmic) while we see a rapid transition to the final value $\lambda_{\kappa=0}\approx\lambda_\Lambda/2$ in the infrared. In both panels, the dashed lines show the corresponding flows in Minkowski space. The Minkowski theory is critical in that case: the longitudinal and transverse mass vanish at $\kappa=0$.}
\label{largeNflowcritical}
\end{center}
\end{figure}
\begin{figure}[t]
\begin{center}
\includegraphics[width=0.4\textwidth]{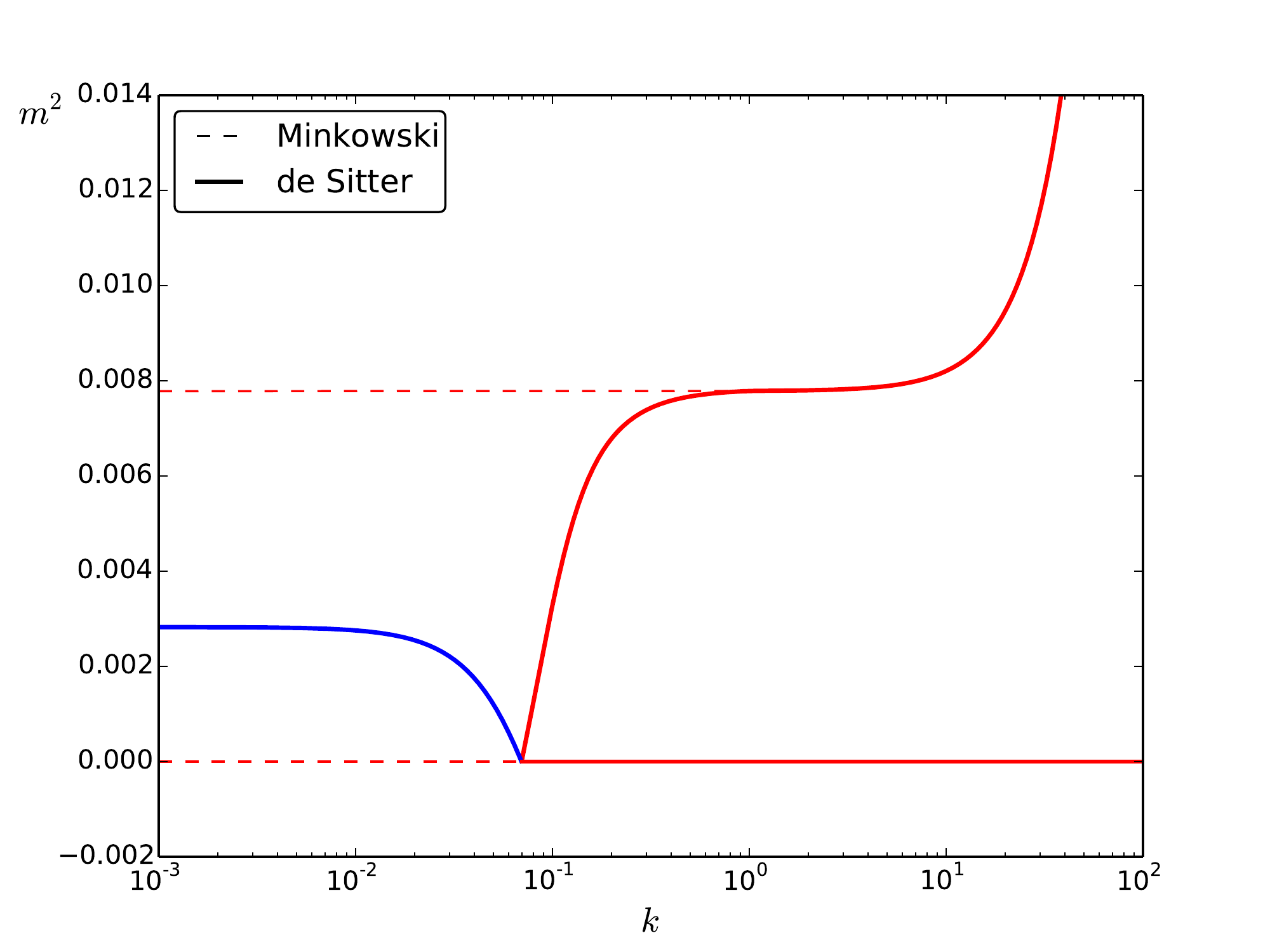}\\
\vspace{.3cm}
\includegraphics[width=0.4\textwidth]{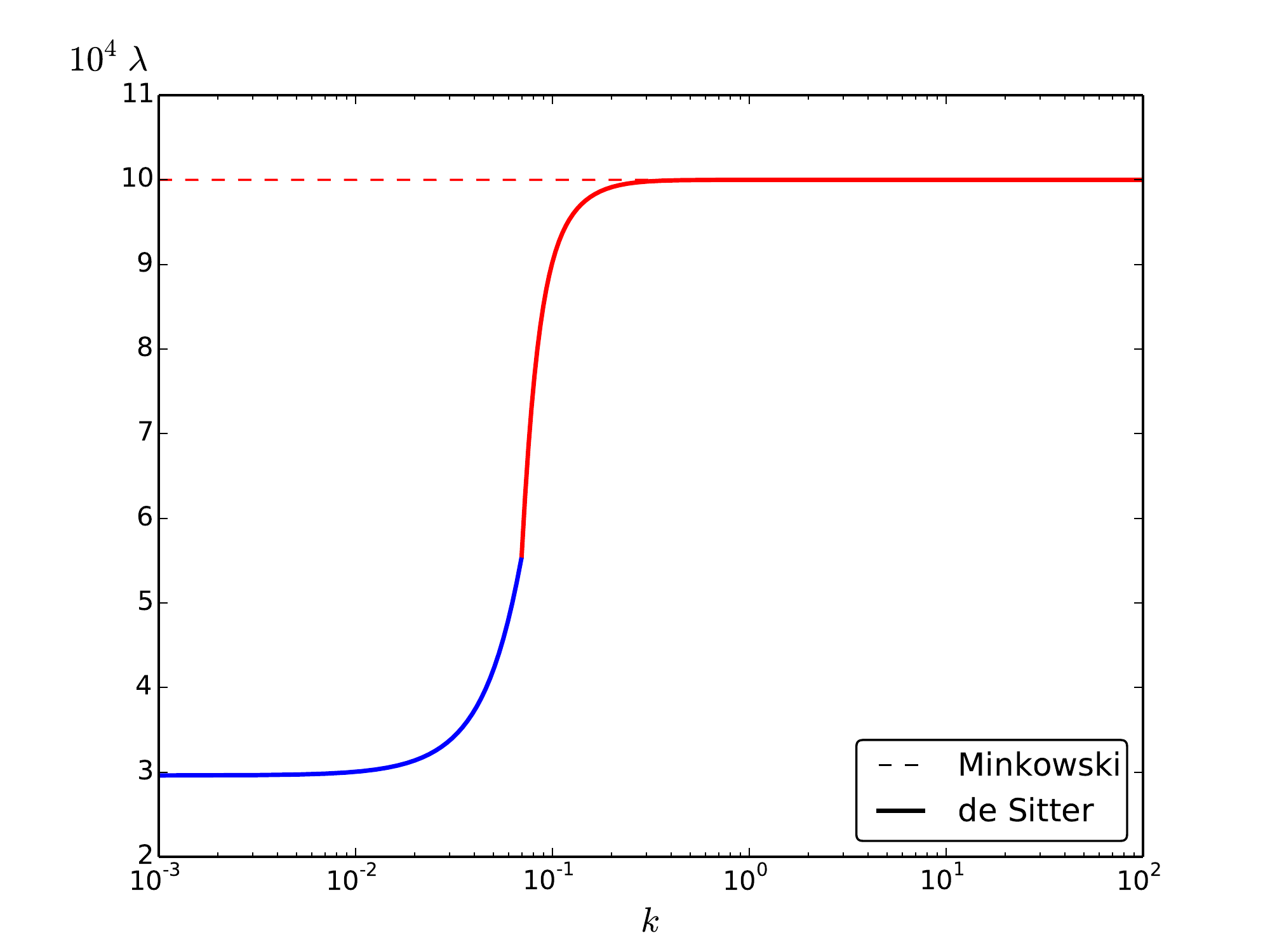}
\caption{Same as in \Fig{largeNflowcritical} for a UV theory in the broken phase. Here, we chose $\bar\rho_\Lambda=25>\rho_c$ and $\lambda_\Lambda=10^{-3}$. The symmetry gets restored deeper in the infrared and the generated mass is thus smaller than in the critical case. The smaller infrared mass implies a smaller infrared coupling as can be seen from \Eqn{eq:finalcoupling}. The dashed lines show the corresponding flows in Minkowski space. We see that the Minkowski theory is in the broken phase at $\kappa=0$, with massless Goldstone modes and a massive longitudinal mode.}
\label{largeNflowbroken}
\end{center}
\end{figure}

The flow of the coupling $\lambda_\kappa$ in the UV regime is given by \Eqn{eq:couplingflow}. In the infrared regime, it can be obtained directly from \Eqn{eq:hahaha} as
\beq
 \frac{1}{\lambda_\kappa}=\frac{1}{\lambda_{\kappa_0}}-\frac{1}{2\Omega_{D+1}}\left\{\frac{1}{\left(\kappa_0^2+m_{t,\kappa}^2\right)^2}-\frac{1}{\left(\kappa^2+m_{t,\kappa}^2\right)^2}\right\}.
\eeq
Alternatively, it can be computed by evaluating the second derivative of the approximate solution \eqn{eq:potflowlargeNIR1} for the potential at the minimum. As recalled above, the  transverse mass is zero as long as $\bar\rho_\kappa\neq0$. Once the symmetry gets restored, the flow of the degenerate mass is obtained from \Eqn{eq:potflowlargeNIR2} as $m_\kappa^2=U_\kappa'(0)=M^2_\kappa(0)-\kappa^2$, that is,
\beq
 m_\kappa^2=\frac{m_{\kappa_0}^2-\kappa^2}{2}+\sqrt{\left(\frac{m_{\kappa_0}^2+\kappa^2}{2}\right)^2+\frac{\lambda_{\kappa_0}}{2\Omega_{D+1}}\left(1-\frac{\kappa^2}{\kappa_0^2}\right)}.
\eeq
In particular, these converge to the final values for $\kappa\to0$
\beq
\label{eq:finalmass}
 m_{\kappa=0}^2=\frac{m_{\kappa_0}^2}{2}+\sqrt{\frac{m_{\kappa_0}^4}{4}+\frac{\lambda_{\kappa_0}}{2\Omega_{D+1}}}
\eeq
and
\beq
\label{eq:finalcoupling}
 \lambda_{\kappa=0}=\lambda_{\kappa_0}\left(1+\frac{\lambda_{\kappa_0}}{2\Omega_{D+1}m^4_{\kappa=0}}\right)^{-1}.
\eeq
Equation \eqn{eq:finalmass} reproduces the result of Ref.~\cite{Serreau:2011fu}. The nonanalytic expression of the generated mass and coupling at the scale $\kappa=0$ in terms of the coupling $\lambda_{\kappa_0}$ is a signature of the nontrivial infrared physics at work here.

Two cases are of interest. The first one is that of a theory which would be close to critical in Minkowski space, i.e., $\bar\rho_\Lambda\approx\rho_c$. In that case, the symmetry gets almost restored already at the horizon scale and the whole infrared flow takes place in the restored symmetry phase. The (dimensionless) effective coupling of the zero-dimensional theory is large, $\lambda_{\kappa_0}^{\rm eff}\equiv\lambda_{\kappa_0}/(2\Omega_{D+1}m_{\kappa_0}^4)\gg1$ and the infrared generated mass and coupling are given by 
\beq
 m_{\kappa=0}^2\approx\sqrt{\frac{\lambda_{\kappa_0}}{2\Omega_{D+1}}}\quad{\rm and}\quad
 \lambda_{\kappa=0}\approx\frac{\lambda_{\kappa_0}}{2}.
\eeq
This reproduces the result of the stochastic approach in the large-$N$ limit for the so-called dynamical mass \cite{Gautier:2015pca}. We note that the dimensionally reduced infrared theory is strongly coupled: 
\beq
 \lambda_{\kappa=0}^{\rm eff}=\frac{\lambda_{\kappa=0}}{2\Omega_{D+1}m^4_{\kappa=0}}\approx\frac{1}{2}.
\eeq

The other interesting limit is that of a theory which would be deeply in the broken phase in Minkowski space ($\bar\rho_\Lambda\gg\rho_c$). In that case, part of the infrared de Sitter flow takes place in the broken phase and the symmetry gets restored in the deep infrared. There remains less RG time to build up a mass and the latter is thus smaller than in the previous critical case. Here, one has $m_{\kappa_0}^2<0$ and, in the limit where $\lambda_{\kappa_0}^{\rm eff}\ll1$, we obtain, for the infrared mass and coupling,
\beq
 m_{\kappa=0}^2\approx \lambda_{\kappa_0}^{\rm eff}|m_{\kappa_0}^2|\quad{\rm and}\quad
 \lambda_{\kappa=0}\approx \lambda_{\kappa_0}^{\rm eff}\lambda_{\kappa_0}.
\eeq
We note that despite the fact that the effective coupling at the horizon scale $\lambda_{\kappa_0}^{\rm eff}\ll1$, the resulting zero-dimensional theory is, again, strongly coupled in the deep infrared:
\beq
  \lambda_{\kappa=0}^{\rm eff}=\frac{\lambda_{\kappa=0}}{2\Omega_{D+1}m^4_{\kappa=0}}\approx1.
\eeq

We show in \Fig{largeNflowcritical} the flow of the longitudinal and transverse masses as well as that of the coupling for the would-be critical theory in $D=3+1$. The case of a theory in the would-be broken phase is shown in \Fig{largeNflowbroken}.

\section{Finite $N$}
\label{sec:finiteN}

We now discuss the flow equation \eqn{eq:flot_N} for $N$ finite. The longitudinal mode plays an increasingly important role as $N$ decreases down to $N=1$, where there are no transverse modes left. As already discussed, nontrivial gravitational effects occur when the local curvature of the potential at the horizon scale $\kappa_0\sim1$ is small, namely, $m_{l,\kappa_0}^2(\rho)\lesssim\kappa_0^2$ and/or $m_{t,\kappa_0}^2(\rho)\lesssim\kappa_0^2$. This is the case for theories which are close to critical or in the broken phase in the UV sense (i.e., theories which would flow toward a critical theory or a broken phase in Minkowski space). For $N\ge2$ the condition of small potential curvature in the broken phase is guaranteed by the presence of Goldstone modes, for which $m_{t,\kappa}^2=U_\kappa'(\bar\rho_\kappa)=0$. 

\begin{figure}[t]
\begin{center}
\includegraphics[width=0.4\textwidth,]{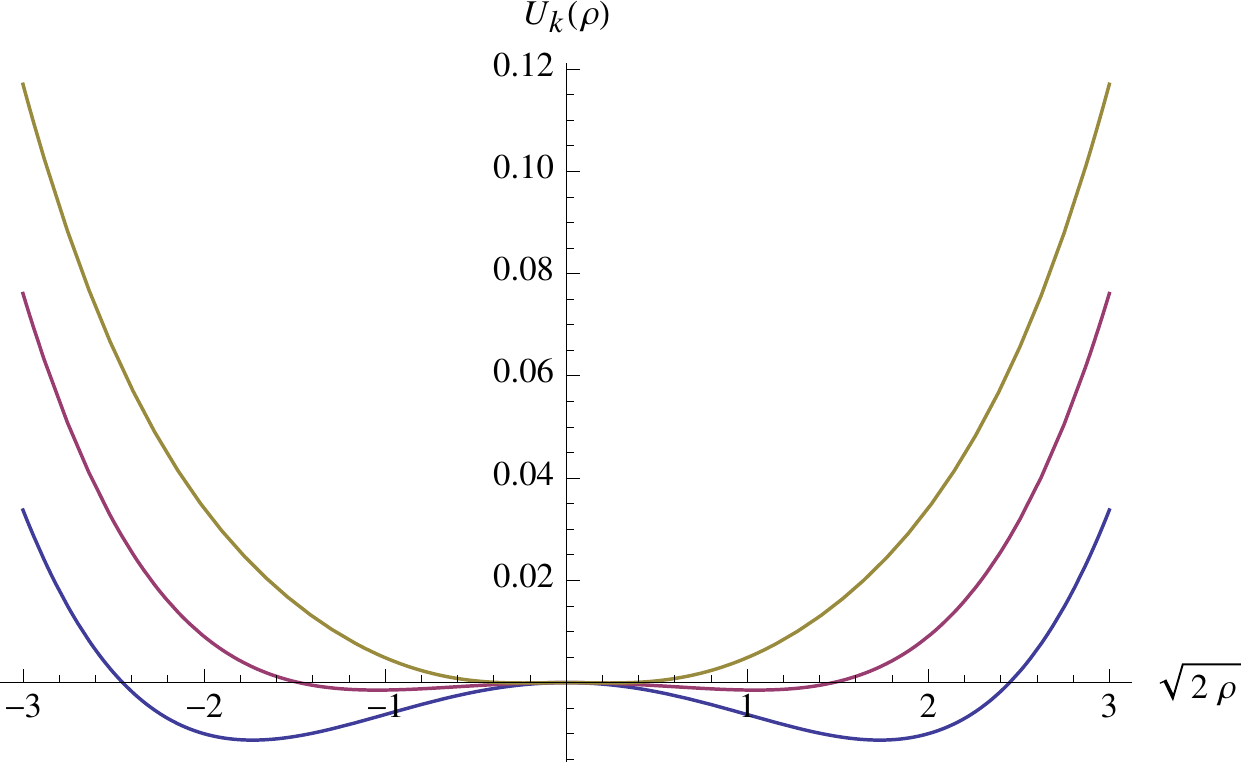}\\
\caption{The effective potential for the $N=1$ theory in $D=3+1$ obtained from the complete functional flow equation \eqn{pot_flow} with initial condition $U_\Lambda(\rho)=\lambda_\Lambda(\rho-\bar\rho_\Lambda)^2/2$ at the ultraviolet scale $\Lambda=10$, with $\lambda_\Lambda=0.01$ and $\bar\rho_\Lambda=1.5$. Curves from bottom to top correspond to $\kappa=10,1,0.1$ One clearly observes the convexification of the potential in the Minkowski regime $\kappa\gtrsim1$ and the symmetry restoration in the infrared regime $\kappa\lesssim1$.}
\label{flow_pot_N1}
\end{center}
\end{figure}

However, there is another mechanism which drives the system into the interesting infrared regime, namely the convexification of the potential along the flow \cite{Berges:2000ew,Delamotte:2007pf}. This simply stems from the fact that, if the theory is properly regulated, one has $\kappa^2+m_{l,\kappa}^2(\rho)>0$ and $\kappa^2+m_{t,\kappa}^2(\rho)>0$ for all scales. In particular, starting the flow in the broken phase at a given ultraviolet scale, the inner region of negative potential curvature between the potential minima is brought to a nearly flat profile at the horizon scale, with a (negative) curvature at most of the order of $\kappa_0^2$. This is a sufficient condition for the flow at superhorizon scales to enter the dimensionally reduced regime mentioned above. For $N=1$, this second, convexification mechanism is the only one at work. This is illustrated in \Fig{flow_pot_N1}, where we show the convexification of the potential\footnote{A qualitative way to understand this convexification effect is to note that the beta function for the potential is positive and is a decreasing function of the curvature $V''_\kappa$. It follows that the overall potential decreases along the flow and that the smaller the curvature, the quicker the flow. The overall effect is to flatten regions of negative curvature. We mention though that, for some initial conditions, this effect is not strong enough and the flow reaches the singular point $\kappa^2+V_\kappa''=0$. This has also been observed in flat space and is a mere artifact of the infrared regulator \cite{Berges:2000ew}. This is usually avoided by using a more appropriate function $R_\kappa(p)$.} along the flow in the UV regime and the subsequent symmetry restoration (complete convexification) due to the effective dimensional reduction in the infrared regime.

We conclude that the qualitative discussion of the large-$N$ case goes over to finite $N$: for initial conditions corresponding to the would-be critical or broken phase cases, the flow enters the dimensionally reduced regime in the infrared. It follows that the symmetry gets restored at a finite RG scale and that a nonzero mass is generated. The latter can be exactly computed from the equivalent integral \eqn{eq:ordinaryint}; see Appendix~\ref{appsec:IRflowequiv}. As before, we parametrize the effective potential at the horizon scale as $U_{\kappa_0}(\rho)=m_{\kappa_0}^2\rho+\lambda_{\kappa_0}\rho^2/2$ and we define $\lambda_{\kappa_0}^{\rm eff}=\lambda_{\kappa_0}/(2\Omega_{D+1}m_{\kappa_0}^4)$. For the critical case ($m_{\kappa_0}^2\approx0$ and $\lambda_{\kappa_0}^{\rm eff}\gg1$), we get
\beq
\label{eq:m2sym}
 m_{\kappa=0}^2={\cal A}(N)\sqrt{\frac{\lambda_{\kappa_0}}{2\Omega_{D+1}}}
\eeq
and
\beq
 \frac{\lambda_{\kappa=0}}{\lambda_{\kappa_0}}=\frac{N{\cal A}^2(N)}{2}\left(1-\frac{{\cal A}^2(N)}{1+2/N}\right),
\eeq
where we defined\footnote{The large-$N$ results of the previous section are recovered using ${\cal A}(N)=1+1/(2N)+{\cal O}(N^{-2})$.}
\beq
\label{eq:AN}
 {\cal A}(N)=\frac{\sqrt{N}}{2}\frac{\Gamma\left(\frac{N}{4}\right)}{\Gamma\left(\frac{N+2}{4}\right)}.
\eeq
In that case, the effective coupling of the dimensionally reduced theory in the infrared is
\beq
 \lambda_{\kappa=0}^{\rm eff}=\frac{N}{2}\left(1-\frac{{\cal A}^2(N)}{1+2/N}\right)>0.135.
\eeq

In the broken symmetry case ($m_{\kappa_0}^2<0$ and $\lambda_{\kappa_0}^{\rm eff}\ll1$), we obtain
\beq
\label{eq:folllow}
 m_{\kappa=0}^2\approx \lambda_{\kappa_0}^{\rm eff}|m_{\kappa_0}^2|\quad{\rm and}\quad
 \lambda_{\kappa=0}\approx \frac{N}{N+2}\lambda_{\kappa_0}^{\rm eff}\lambda_{\kappa_0}
\eeq
and the effective coupling is
\beq
 \lambda_{\kappa=0}^{\rm eff}=\frac{N}{N+2}>\frac{1}{3}.
\eeq

\section{Conclusion}

We have studied the RG flow of O($N$) scalar theories in de Sitter space-time by means of NPRG techniques with particular emphasis on the onset of gravitational effects as one progressively integrates out degrees of freedom from subhorizon to superhorizon momentum scales. At the level of the effective potential, the gravitational enhancement of superhorizon fluctuations results in an effective dimensional reduction of the original $D$-dimensional Lorentzian action to an effective zero-dimensional Euclidean theory. The latter is equivalent to the late-time equilibrium state of the stochastic approach and to the nonperturbative description of the zero mode on the compact Euclidean de Sitter space. The phenomenon of dimensional reduction thus provides a unifying description of these two approaches and explains their identical results for what concerns the calculation of the effective potential.

The present NPRG approach offers a new perspective on the nonperturbative dynamics of light scalar fields on de Sitter space-time. The LPA can be systematically improved, e.g., by employing a derivative expansion \cite{Delamotte:2007pf} or by means of more advanced approximation schemes such as that put forward in Ref.~\cite{Blaizot:2005xy}. This might open a new way for practical calculations of correlation functions of interacting fields in de Sitter space-time. For instance, it is interesting to investigate the role of the field anomalous dimension on the RG flow and to make link with the recent calculation of field correlators at unequal space-time points of Ref.~\cite{Gautier:2013aoa}. This is work in progress.

Other interesting extensions of the present work concern the application of the NPRG approach to other degrees of freedom, such as fermionic or gauge fields, as well as to other types of (e.g., derivative) interactions, for which a stochastic description is not always available \cite{Tsamis:2005hd,Miao:2006pn,Prokopec:2007ak}. An important example is the case of gravitational fluctuations. Finally, it is of interest to investigate the possible implications of the dimensional reduction discussed here for the phenomenology of inflationary cosmology or for models of dark energy.

\section*{Acknowledgements}

We are grateful to D. Benedetti, B. Delamotte, M. Tissier, and N. Wschebor for useful discussions and helpful suggestions. We also thank B. Delamotte and N. Wschebor for useful remarks concerning the manuscript.

\appendix

\section{Flow in Minkowski space}
\label{appsec:Mink}

We derive the LPA flow equation for the effective potential in Minkowski space-time using the regulator (on the closed time contour) given by Eqs.~\eqn{eq:regreg} and \eqn{reg}. Following the procedure outlined in \Sec{sec:setup}, we get, for $N=1$ and leaving the field dependence implicit,
\beq
 \dot V_\kappa = \frac{1}{2} \int \frac{d^dp}{(2\pi)^d}  \dot R_\kappa(p) \big|\chi_\kappa(p,t)\big|^2,
\eeq
where the mode function $\chi_\kappa$ is now defined by
\begin{equation}
\left(\partial_t^2 + p^2 +R_\kappa(p)+ V_\kappa''\right)\chi_\kappa(p,t) = 0.
\end{equation}
With the regulator \eqn{reg} and selecting positive frequency solutions in the infinite past---corresponding to the Minkowski vacuum---we get
\begin{align}
\chi_\kappa(p,t) &= \frac{e^{-i\omega_\kappa(\kappa) t}}{\sqrt{2\omega_\kappa(\kappa)}}\quad{\rm for} \quad p\le \kappa\\
\chi_\kappa(p,t) &= \frac{e^{-i\omega_\kappa(p) t}}{\sqrt{2\omega_\kappa(p)}}\quad{\rm for} \quad p \ge \kappa,
\end{align}
with $\omega_\kappa(p) = \sqrt{p^2 + V_\kappa''}$. Using $\dot R_\kappa(p)=2\kappa^2\theta(\kappa^2-p^2)$, the Minkowski flow equation thus reads
\beq
\label{appeq:flowMink}
\dot V_\kappa =\frac{\Omega_d }{2d(2\pi)^d} \frac{\kappa^{d+2}}{\sqrt{\kappa^2+V''_\kappa}},
\eeq
which agrees with \Eqn{eq:betaMink}.
The generalization to $N\ge1$ is straightforward; see Eqs.~\eqn{eq:flot_N} and \eqn{eq:flot_N2}.

It is a simple exercise to show that the flow equation \eqn{appeq:flowMink} reproduces the standard one-loop results for the critical exponents of O($N$) models in $D=4-\epsilon$ dimensions. To this aim it is sufficient to consider the polynomial ansatz
\begin{equation}
U_\kappa(\rho) = \frac{\lambda_\kappa}{2} \left (\rho-\bar \rho_\kappa \right )^2. \label{ansatz}	
\end{equation}
The parameters  $\bar \rho_\kappa$ and $\lambda_\kappa$ are defined as 
\begin{align}
U_\kappa'( \bar \rho_\kappa) = 0\,, \quad U_\kappa''( \bar \rho_\kappa) = \lambda_\kappa 
\end{align}
and satisfy the following flow equations
\begin{align}
\dot{\bar \rho}_\kappa &=\frac{v_d}{2N}\kappa^{d+2}\left[ \dfrac{3 }{(\kappa^2 + 2\lambda_\kappa \bar\rho_\kappa)^{\frac{3}{2}}} + \dfrac{N-1}{\kappa^3} \right] \\
 \dot \lambda_\kappa &=\frac{3v_d}{4N}\kappa^{d+2} \lambda_\kappa^2 \left[ \dfrac{9}{(\kappa^2 + 2\lambda_\kappa \bar\rho_\kappa)^{\frac{5}{2}}}  +\dfrac{N-1}{\kappa^5}\right],
 \end{align} 
where $v_d=\Omega_d/[2d(2\pi)^d]$. Introducing the dimensionless parameters
\begin{equation}
 r _\kappa = \bar \rho_\kappa \kappa^{2-D} \quad{\rm and}\quad \ell_\kappa = \lambda_\kappa \kappa^{D-4}
\end{equation}
and expanding to first nontrivial order in $\ell_\kappa\sim{\cal O}(\epsilon)$ close to the Wilson-Fisher fixed point, we have
\begin{align}
\dot r_\kappa &= -\left(2-\epsilon+\frac{9v_d}{2N}\ell_\kappa\right) r_\kappa +\frac{v_d(N+2)}{2N}+{\cal O}(\epsilon^2)\\ 
\dot \ell_\kappa &= -\epsilon\ell_\kappa + \frac{3v_d(N+8)}{4N}  \ell_\kappa^2 + {\cal O}(\epsilon^3).
\end{align}
The fixed point is located at
\begin{equation}
r^* = \frac{v_d(N+2)}{4N} \quad{\rm and}\quad\ell^* = \frac{4 N\epsilon}{3 v_d(N+8)}.  
\end{equation}
Critical exponents are obtained from the linearized flow around the fixed point. For instance, the correlation-length exponent $\nu$ is obtained as minus the inverse of the smallest (negative) eigenvalue of the Jacobian matrix of the linearized flow \cite{Delamotte:2007pf}. We get
\begin{equation}
\nu = \frac{1}{2} + \frac{\epsilon}{4}\frac{N+2}{N+8} + O(\epsilon^2),
\end{equation}
which reproduces the well-known perturbative result \cite{ZinnJustin:2002ru}.

\section{Dimensionally reduced RG flow}
\label{appsec:IRflowequiv}

In this section we show how the flow of the parameters describing the effective potential in the regime of dimensional reduction can be read off the equivalent zero-dimensional theory, \Eqn{eq:ordinaryint}. For the sake of the discussion we focus on the symmetric phase and we only consider the square mass and the quartic coupling, defined as
\beq
 m_\kappa^2=U_\kappa'(0)\quad{\rm and}\quad\lambda_\kappa=U_\kappa''(0).
\eeq
The discussion can easily be extended to any other coupling. At vanishing sources, the first nontrivial correlators have the following O($N$) structures
\beq
 \langle\varphi_a\varphi_b\rangle=\delta_{ab}G_\kappa
\eeq
and
\beq
 \langle\varphi_a\varphi_b\varphi_c\varphi_d\rangle=(\delta_{ab}\delta_{cd}+\delta_{ac}\delta_{bd}+\delta_{ad}\delta_{bc})C_\kappa^{(4)}.
\eeq
The two- and four-point functions $G_\kappa$ and $C_\kappa^{(4)}$ are related to the parameters of the effective potential $U_\kappa(\rho)$ through the Legendre transform \eqn{eq:legendre} as
\beq
\label{appeq:ee}
 G_\kappa=\frac{1}{\Omega_{D+1}(\kappa^2+m_\kappa^2)}
\eeq
and
\beq
 C^{(4)}_\kappa=G_\kappa^2-\frac{\Omega_{D+1}\lambda_\kappa}{N}G_\kappa^4.
\eeq

For a potential at the horizon scale of the form $U_{\kappa_0}(\rho)\approx m_{\kappa_0}^2\rho+\lambda_{\kappa_0}\rho^2/2$, the various correlators of the theory are obtained from the moments
\beq
\label{appeq:moments}
 \langle\left(\varphi_a\varphi_a\right)^q\rangle=\frac{\int_0^\infty d\varphi\varphi^{N+2q-1}e^{-\alpha\varphi^2-\beta\varphi^4}}{\int_0^\infty d\varphi\varphi^{N-1}e^{-\alpha\varphi^2-\beta\varphi^4}},
\eeq
where we introduced $\alpha=\Omega_{D+1}(\kappa^2+m_{\kappa_0}^2)/2$ and $\beta=\Omega_{D+1}\lambda_{\kappa_0}/(8N)$. For instance, one has $G_\kappa=\langle\varphi_a\varphi_a\rangle/N$ and $C_\kappa^{(4)}=\langle(\varphi_a\varphi_a)^2\rangle/[N(N+2)]$. The moments \eqn{appeq:moments} can easily be computed. For instance, in the limit $\beta/\alpha^2\gg1$, which corresponds to the critical case discussed in the main text, one has
\beq
\label{appeq:rr}
 \langle\left(\varphi_a\varphi_a\right)^q\rangle\approx\beta^{-{q\over2}}\frac{\Gamma\left(\frac{N+2q}{4}\right)}{\Gamma\left(\frac{N}{4}\right)}.
\eeq
Putting Eqs.~\eqn{appeq:ee}--\eqn{appeq:rr} together, one obtains Eqs.~\eqn{eq:m2sym}--\eqn{eq:AN}.
The other limit of interest is that of a would-be broken phase, corresponding to $\alpha<0$ and $\beta/\alpha^2\ll1$. In that case, one gets
\beq
 \langle\left(\varphi_a\varphi_a\right)^q\rangle\approx\left(\frac{|\alpha|}{2\beta}\right)^q,
\eeq
from which \Eqn{eq:folllow} follows.

\end{document}